\documentclass[longauth]{aa} 
\usepackage{graphicx}
\usepackage{txfonts}
\begin{document} 
   \title{The Gaia-ESO Survey: the origin and evolution of s-process elements 
   \thanks{Based on observations collected with the FLAMES instrument at
VLT/UT2 telescope (Paranal Observatory, ESO, Chile), for the Gaia-
ESO Large Public Spectroscopic Survey (188.B-3002, 193.B-0936).}}

   \author{L. Magrini\inst{1}, L.Spina \inst{2}, S. Randich\inst{1}, E. Friel, \inst{3}, G. Kordopatis\inst{4}, C. Worley\inst{5}, E. Pancino\inst{1,6}, A. Bragaglia\inst{7}, P. Donati\inst{7},  G. Tautvai\v{s}ien\.{e}\inst{8}, 
   V. Bagdonas\inst{8}, E. Delgado-Mena\inst{9}, V. Adibekyan\inst{9}, S. G. Sousa\inst{9}, 
   F. M. Jim\'{e}nez-Esteban\inst{10}, N. Sanna\inst{1}, V. Roccatagliata\inst{1}, R. Bonito\inst{11}, L. Sbordone\inst{12}, S. Duffau\inst{13}, 
   G. Gilmore\inst{5},  S. Feltzing\inst{14},  R.~D. Jeffries\inst{15},  A. Vallenari\inst{16}, E.~J. Alfaro\inst{17}. T. Bensby\inst{14},  P. Francois\inst{18}, S. Koposov\inst{5}, A.~J. Korn\inst{19}, A. Recio-Blanco\inst{20},   R. Smiljanic\inst{21},  A. Bayo\inst{22,23},  G. Carraro\inst{24}, A.~R. Casey\inst{2},  M.~T. Costado\inst{25}, 
   F. Damiani\inst{11},  E. Franciosini\inst{1},  A. Frasca\inst{26}, A. Hourihane\inst{5}, P. Jofr\'e \inst{27}, P. de Laverny\inst{20},  J. Lewis\inst{5},   T. Masseron\inst{28,29},  L. Monaco\inst{13}, L. Morbidelli\inst{1}, L. Prisinzano\inst{11}, G. Sacco\inst{1}, S. Zaggia\inst{16}}
 \institute{INAF - Osservatorio Astrofisico di Arcetri, Largo E. Fermi, 5, I-50125 Firenze, Italy
\email{laura@arcetri.astro.it} \and 
Monash Centre for Astrophysics, School of Physics and Astronomy, Monash University, VIC 3800, Australia \and
Department of Astronomy, Indiana University, Bloomington, IN, USA\and
Universit\'e C\^{o}te d'Azur, Observatoire de la C\^{o}te d'Azur, CNRS, Laboratoire Lagrange, France  \and
Institute of Astronomy, Madingley Road, University of Cambridge, CB3 0HA, UK\and
Space Science Data Center - Agenzia Spaziale Italiana, Via del Politecnico SNC, 00133 Roma\and
INAF-Osservatorio di Astrofisica e Scienza dello Spazio, via Gobetti 93/3, I-40129 Bologna, Italy \and
Astronomical Observatory, Institute of Theoretical Physics and Astronomy, Vilnius University, Saul\.{e}tekio al. 3, LT-10222 Vilnius, Lithuania\and
Instituto de Astrofisica e ciencias do espa\c co -CAUP, Universidade do Porto, Rua das Estrelas, P-4150-762 Porto, Portugal\and
Departamento de Astrof\'{\i}sica, Centro de Astrobiolog\'{\i}a (INTA-CSIC), ESAC Campus, Camino Bajo del Castillo s/n, E-28692 Villanueva de la Ca\~nada, Madrid, Spain\and
INAF - Osservatorio Astronomico di Palermo, Piazza del Parlamento 1, 90134, Palermo, Italy\and
European Southern Observatory, Alonso de Cordova 3107 Vitacura, Santiago de Chile, Chile\and
Departamento de Ciencias Fisicas, Universidad Andres Bello, Fernandez Concha 700, Las Condes, Santiago, Chile\and
Lund Observatory, Department of Astronomy and Theoretical Physics, Box 43, SE-221 00 Lund, Sweden\and
Astrophysics Group, Keele University, Keele, Staffordshire ST5 5BG, United Kingdom   \and
INAF - Padova Observatory, Vicolo dell'Osservatorio 5, 35122 Padova, Italy\and
Instituto de Astrof\'{i}sica de Andaluc\'{i}a-CSIC, Apdo. 3004, 18080, Granada, Spain\and
GEPI, Observatoire de Paris, CNRS, Universit\'e Paris Diderot, 5 Place Jules Janssen, 92190 Meudon, France\and
Department of Physics and Astronomy, Uppsala University, Box 516, SE-751 20 Uppsala, Sweden\and
Laboratoire Lagrange (UMR7293), Universit\'e de Nice Sophia Antipolis, CNRS,Observatoire de la C\^ote d'Azur, CS 34229,F-06304 Nice cedex 4, France\and
Nicolaus Copernicus Astronomical Center, Polish Academy of Sciences, ul. Bartycka 18, 00-716, Warsaw, Poland\and
Instituto de F\'isica y Astronomi\'ia, Universidad de Valparai\'iso, Chile\and
Millennium Nucleus ``N\'ucleo Planet Formation", Universidad de Valpara\'iso\and
Dipartimento di Fisica e Astronomia, Universit\`a di Padova, Vicolo dell'Osservatorio 3, 35122 Padova, Italy\and
Departamento de Did\'actica, Universidad de C\'adiz, 11519 Puerto Real, C\'adiz, Spain\and
INAF - Osservatorio Astrofisico di Catania, via S. Sofia 78, 95123, Catania, Italy\and
N\'ucleo de Astronom\'{i}a, Facultad de Ingenier\'{i}a, Universidad Diego Portales, Av. Ej\'ercito 441, Santiago, Chile\and
Instituto de Astrof\'{\i}sica de Canarias, E-38205 La Laguna, Tenerife, Spain\and        
Universidad de La Laguna, Dept. Astrof\'{\i}sica, E-38206 La Laguna, Tenerife, Spain
}
    \date{Received ; accepted }

 
  \abstract
   {Several works have found an increase of the abundances of the s-process neutron-capture elements in the youngest Galactic stellar populations. These trends give important constraints to stellar and Galactic evolution and they need to be confirmed 
   with large and statistically significant samples of stars spanning wide age and distance intervals.  }
   {We aim to trace the abundance patterns and the time-evolution of  five s-process elements - two belonging to the first peak, Y and Zr, and three belonging to the second peak, Ba, La and Ce - using  the Gaia-ESO {\sc idr5} results for open clusters and 
   disc stars. }
   {From the UVES spectra of cluster member stars, we determined the average composition of clusters with ages $>$0.1~Gyr.  
   We derived statistical ages and distances of field stars, and we separated them in thin and thick disc populations. 
   We studied the time-evolution and dependence on metallicity of abundance ratios using open clusters and field stars whose parameters and abundances were derived in a  homogeneous way.  }
   {Using our large and homogeneous sample of open clusters, thin and thick disc stars, spanning an age range larger than 10~Gyr, we confirm an increase towards young ages of s-process abundances in the Solar neighbourhood. 
   These trends are well defined for open clusters and stars located nearby the solar position and they may be explained by a late enrichment due to significant contribution to the production of these elements from long-living low-mass stars. 
   At the same time, we found a strong dependence of the s-process abundance ratios with the Galactocentric distance and with the metallicity of the clusters and field stars. 
    }
   {Our results, derived  from the largest and homogeneous sample of s-process abundances in the literature, confirm the growth with decreasing stellar ages of the s-process abundances in both field and open cluster stars.  At the same time, taking advantage of the abundances of open clusters located in 
   a wide Galactocentric range,  they open a new view on the dependence of the s-process evolution on the metallicity and star formation history, pointing to different behaviours at various Galactocentric distances. }

\authorrunning{Magrini, L. et al.}
\titlerunning{S-process in Open Clusters and Galactic discs }

   \maketitle
%

\section{Introduction}

Elemental abundances and abundance ratios in the different populations of our  Galaxy provide fundamental constraints to the scenarios 
of galaxy formation and evolution. 
With a large number of different elements  we can study the different processes and time scales 
involved in stellar and galactic evolution. 

The heavy elements (atomic number Z$>$30) are produced by subsequent capture of neutrons on lighter elements. Their evolution in the Galaxy has been studied since the pioneering work of \citet{PT97, travaglio99}. 
There are 54 stable or long-lived neutron-capture elements, compared to only 30 lighter elements \citep[cf.][]{sneden08}. However, they are much less abundant than the lighter elements and correspond to an abundance by number of  10$^{-8}$ in the Sun. 
Their production is an endothermic process. In addition, the Coulomb barriers increase with the proton number. Consequently, the nuclei heavier than iron 
can be created only with a series of neutron capture events:  for those events there are no Coulomb barriers. 
Neutron-capture elements can be split into two different groups: those formed mainly  via slow neutron-capture ``s-process'' and those produced by rapid neutron capture ``r-process''. The terms slow (time-scale for a single neutron capture of hundreds or thousands of years) or rapid (fractions of second) are so defined 
in comparison with the time-scales of the $\beta$-decay of the nuclei on which neutrons are accreted.  For the s-process  and r-process,   low, i.e. 10$^7$ to 10$^{11}$ cm$^{-2}$ s$^{-1}$, and high fluxes of neutrons, i.e. $\sim$10$^{20}$ cm$^{-2}$ s$^{-1}$,  are needed, respectively.

The s-process dominated elements, hereafter s-process elements,  are distributed primarily in two peaks in the periodic table around the neutron magic  numbers  N = 50, 82.
The first s-process-peak is located around N=50 and leads to the formation of Sr, Y, and Zr (light-s elements); the second peak 
at N=82 produces Ba, La, Ce, Pr, and Nd. There is also a  third peak that corresponds to the production of Pb at  N=126.
 The s-process can be further divided into a ``main'' process, a ``weak'' process and a ``strong" process: 
the first occurs during the  asymptotic giant branch (AGB) phases \citep{gallino98, busso99}, whereas the second takes place in  massive stars that produce elements with N$\leq$88 \citep{raiteri93, pignatari10} and the third is responsible for about 50\% of solar $^{208}$Pb production by low-metallicity AGB stars \citep[e.g.][]{gallino98}. 

Recently, several works found an increase of the abundances of the s-process elements for the youngest stellar populations. 
Among the first studies that noticed it, \citet{dorazi09} reported the discovery of increasing barium abundance with decreasing age for a large sample of Galactic open clusters. 
They gave a first tentative explanation assuming  a higher Ba yield from the s-process in low-mass stars than the average suggested by parameterised models of neutron-capture nucleosynthesis. However, their chemical evolution model was not able to explain the very high [Ba/Fe] measured in the youngest open clusters of their sample. Other studies followed, all finding strong trends of Ba abundance with age.

A number of studies \citep[e.g.,][]{maiorca11, JF13, mish13, mish15} extended the analysis to other s-process elements, both lighter (Y, Zr) and heavier (Ba, La, Ce) neutron-rich nuclei.  Conclusions varied with the studies, painting a somewhat confused picture of age dependencies in elements other than Ba.   \citet{maiorca11} found the elements Y, Ce, Zr, and La to show increasing abundance ratios with decreasing age, as with Ba, but to a lesser degree.  \citet{JF13} found strong age trends in Ba but no age trends in Zr and La, and noted that the conclusions drawn by different studies depended on the sample size and the range of ages sampled, and were affected by sometimes large systematic differences between studies \citep[e.g.][]{yong12}.  \citet{mish13, mish15} found clusters and field stars shared similar behavior in Y and La, with slight, if any, trends with age.  A number of studies found overall enhancements in abundance ratios in the clusters relative to solar, up to 0.2~dex, particularly for [Zr/Fe].

Clusters in these samples not only spanned large age ranges, from several hundred million years to $\sim$8~Gyr, they also covered large distance ranges, in particular sampling the outer disk.  Although open clusters have been shown not to follow any relationship between age and metallicity, they do exhibit gradients in abundance, and their location (at birth) is more of a factor in determining their abundance than is their age.  The interplay of the effects of both age and location introduce complexity into the interpretation of abundance trends, while at the same time offering the potential to illuminate them.  While these cluster studies benefitted from the accurate age determinations of open clusters, particularly their ability to trace the recent epoch, they nevertheless revealed that differing sample sizes, age, and distance distributions, and the non-uniformity of analysis between studies complicated the conclusions that could be drawn, particularly about weak trends. 


Other groups investigated this phenomenon using field stars, especially solar twins. We recall that the ages of field stars can be determined with lower accuracy than those for open clusters at least in the pre-Gaia era. 
The sample of solar-type stars of \citet{dasilva12} showed a pattern of [s/Fe] with age similar to that of open clusters  \citep{dorazi09, maiorca12},  with increasing abundance ratios with decreasing age, although the form of the dependence varied with element: for Ce and Zr, the increase pertained only to stars younger than the Sun, while for Y and Sr, the abundances increased linearly with decreasing age. 
Recently,  \citet{RL17} measured La, Ce, Nd, and Sm: their Figure 7 suggests that the [X/Fe] values measured for these heavy s-process elements display a smoothly increasing weak trend with decreasing stellar age, with changes of $\sim$0.1~dex from 10~Gyr ago to the present. 
For Ba, they found,  like many others, a much stronger age dependence.  They explain the larger enhancement for Ba in younger stars as due to an overestimation of Ba by standard methods of LTE analysis, finding the Ba enhancements to be strongly correlated with the level of chromospheric activity. If this effect is considered and  the very young stars are excluded (the chromospheric activity has a striking decline from 1-10~Myr to the age of the Sun), the enhancement of Ba is in line with that of the other s-process elements. 
In the same framework, \citet{nissen17} analysed a sample of ten stars from the Kepler LEGACY \citep{silva17} and found that over the lifetime of the Galactic disc the abundance of Y increases. The trend of [Y/Fe] with age, and an even stronger trend of [Y/Mg], were previously found by e.g., \citet{dasilva12, nissen16, spina16, feltzing17} and it can be explained by production of yttrium via the slow neutron-capture process in low-mass 
AGB stars \citep{maiorca12, KL16}. 
Finally, \citet{spina18}, in their differential analysis of a sample of solar twin stars,  found an increase with decreasing stellar ages of the abundance ratios [X/Fe] of all the s-process elements of the first and second peaks. 
 They found that the rise of Ba with time is consistent with that of the other n-capture elements, considering its higher contribution from the s-process. In all cases, trends of [X/Fe] are weak, with changes of ~0.1 to ~0.2 dex over the lifetime of the disk.

Thus, 
a general consensus of an increasing, though weak, trend in the s- process elements production exists. 
There are still issues regarding the maximum abundance for Ba, which is still debated \citep[see][]{RL17} and the trends for the open clusters, which are affected by positional dependence, radial gradients, and differences in analysis and samples \citep[see, e.g.][]{jacobson11, yong12}.
Taking inspiration from the conclusions of \citet{JF13}, we need a homogeneous analysis of large samples of open clusters together with a sample of field stars that will facilitate the measurement of many n-capture species in clusters and in the field spanning a wide range in age. 

Among the several on-going spectroscopic surveys, the Gaia-ESO Survey \citep[GES, ][]{gilmore12,RG13},  a ESO large public survey, is providing high resolution spectra of 
different stellar populations of our Galaxy using FLAMES@VLT \citep{pasquini02} employing both Giraffe and UVES fibres in Medusa mode.  
It aims at homogeneously determining stellar parameters and abundances for a large sample of stars in the field, in Galactic open clusters, and in calibration samples, including also globular clusters.
In particular, the high resolution spectra obtained with UVES allow the determination of abundances of a large variety of elements including 
many neutron-capture elements: 
two light s-process elements (Y, Zr), three heavy s-process elements (Ba, La, Ce), one r-process
element (Eu), and three elements with significant contributions from both processes (mixed elements-- Mo, Pr, and Nd).
In the present work we investigate the evolution of the group of s-process elements. 
The paper is structured as follows. 
In Section 2 we present the data reduction and analysis and in Section 3 the solar abundance scale. 
In Sections 4  we describe the field and cluster samples, whereas in Section 5 we show the results on the abundance of s-process elements, and  in Section 6 we discuss their time evolution. 
Finally, in Section 7 we give our summary and conclusions.

\section{The data reduction and analysis}
\label{datared}

We use data from the 5$^{th}$ internal data release ({\sc idr5}) of the Gaia-ESO Survey. For this work we use
 only the products bases on the UVES spectra, which have the  higher spectral resolution (R=47000) and wider spectral range. 
They have been reduced and analysed by the Gaia-ESO consortium.
UVES data reduction is carried out using the FLAMES-UVES
ESO public pipeline \citep{modigliani04}. The UVES data reduction process and the determination of  the radial velocities (RVs) 
are described in \citet{sacco14}. 
Various {\sc WGs} contribute to the spectral analysis of different kinds of stars and/or setups: the data discussed in the present paper have been analysed by WG11 which is in charge of the analysis of the UVES spectra of F-G-K spectral type stars both in the field of the Milky Way (MW) and in  intermediate-age and old
clusters and obtained with the UVES red arm standard setting with central wavelength of 520 and 580 nm. These spectra were analysed with the Gaia-ESO multiple pipelines strategy, as described in \citet{smi14}. The results of each pipeline are combined with an updated methodology (Casey et al., in preparation) to define a final set of recommended values of the atmospheric parameters.
The results of WG11 are homogenised using several calibrators e.g., benchmark stars and open/globular clusters
selected as described in \citet{pancino17} and adopted for the homogenisation by WG15 (Hourihane et al. in preparation). 
The recommended parameters and abundances used in the present work are reported in the final {\sc idr5} catalog, which contains the observations obtained before December 2015.

To measure the usually faint absorption lines of neutron capture elements a high signal-to-noise ratio (SNR) moth spectrum is required. 
For this reason, for the field stars we made a selection of the UVES results based on the SNR and on the quality of the parameter determination. 
 The thresholds of our cuts are defined comparing our results with the very high SNR samples of \cite{bensby14} for Ba and of \citet{BB16} for the other elements, 
designed to study the neutron-capture element abundances.  
To obtain similar dispersions as in \cite{bensby14} and in \citet{BB16} for the neutron-capture element 
distributions (at a given metallicity) we need a minimum SNR of 70 for Milky Way field stars. 
In addition, we discard stars whose errors on stellar parameters are: E$_{\rm Teff} >$150~K, E$_{\rm logg} >$0.25~dex, E$_{\rm [Fe/H]} >$0.20~dex and E$_{\xi} >$0.20~km~s$^{-1}$. 
The final sample of field stars fulfilling these requirements includes about  600 objects. The sample of open clusters includes 22 clusters, with a total of 165 member stars.  

For the cluster stars, we did not apply any selection, because their spectra usually are characterised by higher SNR (and thus almost all stars in cluster stars will be included). In addition, we do not consider individual measurements for cluster stars, but we rely on the cluster median abundances, which are derived from several member stars (the membership analysis is discussed in Section \ref{sec_cluster}). We only discard cluster abundances with high standard deviations $>$ 0.3~dex.

\section{Solar abundance scale}

\label{solar}
 \begin{table*}
\begin{center}
\caption{{\sc idr5} solar and M67 abundances.  }
\tiny
\begin{tabular}{llll}
\hline \hline
Element &  Sun ({\sc idr5}) &Sun \citep{grevesse07} &  M67 giants ({\sc idr5}) \\
\hline
{\sc Y}~II   & 2.19$\pm$0.12		        & 2.21$\pm$0.02 & 2.14$\pm$0.01(0.09)  \\
{\sc Z}r~II   &  2.53$\pm$0.13	               &2.58$\pm$0.02  &2.54$\pm$0.03(0.05)\\
{\sc B}a~II & 2.17$\pm$0.06		        &2.17$\pm$0.07  &2.07$\pm$0.03(0.07)\\
{\sc L}a~II &  -                                         &1.13$\pm$0.05  &1.00$\pm$0.02(0.12) \\
{\sc C}e~II &   1.70$\pm$0.11                   &1.70$\pm$0.10  &1.71$\pm$0.01(0.01)\\
{\sc E}u~II & -						 &0.52$\pm$0.06    & 0.42$\pm$0.01(0.04)\\
\hline \hline
\end{tabular}
\label{tab_sun}
\end{center}
\end{table*}

To obtain abundances on the solar scale, we need to define our abundance reference.  In Table~\ref{tab_sun} we show 
three different sets of abundances:  the solar abundances from  {\sc idr5} computed from archive solar spectra; those from  \citet{grevesse07}; and the abundances 
of giant stars in M67. The abundances are in the usual 12+$\log$(X/H) form. 
The cluster M67 is known to have the same composition as the Sun  \citep[e.g.][]{randich06,pasquini, on14, liu16} and thus it is useful to confirm it with the  {\sc idr5} data. Since small effects of stellar diffusion might be present in the abundances
of the dwarf stars of M67 (see, e.g.,  {\"O}nehag et al. 2014 and Bertelli Motta et al., submitted), we only considered the abundances of the giant stars. 
The GES solar abundances for the s-process elements are in very good agreement with  the reference solar abundances from \citet{grevesse07}. 
The average abundances for the three member giant stars in M67 (T$_{\rm eff}$$\sim$4800-4900 and   $\log$~g$\sim$3-3.4) from the {\sc idr5} recommended values 
are given together with its  standard deviation of the mean  and the typical error on each measurement (in parenthesis) (see third column of Table\ref{tab_sun}). 
In the following, we normalise our abundances to the solar abundances (first column), with the exception of  La for which there
 is no solar abundance available and thus we make use of the M67 abundances. 

\section{The stellar samples}
\subsection{The Milky Way field samples}
\label{sec_stars}
The sample discussed here includes field stars observed with UVES 580 that belong 
to the stars in the Milky Way sample, and in particular to the  solar neighbourhood sample (\texttt{GES\_TYPE = }`GE\_MW') and  the inner disc sample  (\texttt{GES\_TYPE = }`GE\_MW\_BL').
The selection functions of these stars are described in \citet{sto16}. 
For these stars, we computed spectroscopic distances as in \citet{magrini17}, by projecting the stellar atmospheric  parameters and (J-K) 2MASS colours on a set of  isochrones \citep{bressan12} spanning the age range between $0.1$ and $13.9$ Gyr (with a uniform step of 0.1 Gyr) and a metallicity range between $-2.3$ and $+0.2$ dex (with a uniform step of 0.1 dex). The projection takes into account the uncertainties on the atmospheric parameters, the likelihood of a  star to belong to a given evolutionary phase, and the line-of-sight extinction (iteratively corrected by the distance) of \citet{schlegel}. The projection therefore not only returns the absolute magnitude in various bands (and hence, eventually, the line-of-sight distance), but also a crude estimation of the stellar ages. The details of the method are described in \citet{kor11}, with the updates of \citet{rc14} and \citet{kordo15}.  Typical errors ranges from 20\% up to more than 100\%, with the bulk of the stars having errors of the order of 30-50\%. 

We have divided the field stars into thin and thick disc stars, using their [$\alpha$/Fe] abundance ratio \footnote{[$\alpha$/Fe] is computed by averaging [Mg/Fe], [Ca/Fe], [Si/Fe] and excluding [O/Fe], which is affected by large errors in dwarf stars} to discriminate the two populations, following the approach of 
  \citet{adibekyan11}. We have used the multi slope curve of  \citet{adibekyan11} to define thin and thick disc stars on the basis of their [$\alpha$/Fe] and [Fe/H]. 
In the following analysis, we have included stars with T$_{\rm eff} <$ 6120 K \citep[to avoid NLTE effects in some neutron-capture element abundances as discussed in][] {bensby14})  and with SNR$>$70, 
and applied the error cuts described in the previous section (see Sec.\ref{datared}) to exclude poor quality results.
The adopted separation between thin and thick disc stars is also presented in the first panels of Figures~\ref{elfeA} and \ref{elfeB}. 
In the following discussion, we include only thin and thick disc stars located in the Solar neighbourhood (6.5~kpc$<$R$_{\rm GC}<$9.5~kpc) to avoid confusion due to the radial dependence of the abundance ratios.

\subsection{The cluster sample} 
\label{sec_cluster}
We consider the sample of open clusters  with ages$>$0.1~Gyr whose parameters and abundances have been delivered  in {\sc idr5}. 
 We do not include younger clusters because their stars are usually more difficult to analyse and they can be affected by chromospheric effects. 
 
For each cluster we have extracted member stars using the information on their radial velocities. 
We have considered as member stars  those within 1.5-$\sigma$ from the cluster systemic velocity (computed using the UVES spectra) 
and we have excluded outliers in metallicity $|$[Fe/H]$_{\rm star}-<$[Fe/H]$>|>$0.1~dex.
Based on stars assigned as members, we have computed 
the median elemental abundances, expressed in the form [X/H]=log(X/H)-log(X/H)$_{\odot}$, which are presented in Table~\ref{tab_clu_par}. 
The uncertainties reported on each abundance are the standard deviation of cluster member abundances. 
We consider in the following analysis only abundance ratios with dispersion lower than 0.3~dex. 

In Table~\ref{tab_clu_par} we summarise the basic properties of the sample open clusters: coordinates, ages, Galactocentric distances, heights above  the plane, mean radial velocities of cluster members, median metallicity and the number of cluster member stars used to compute the metallicity. 
There is a general agreement with the previous official release \citep[see, e.g.][]{magrini17}, but for some clusters  the median metallicity might have  changed slightly.  
For clusters in common with \citet{magrini17}, we  adopt  the same ages and distances, for the clusters which are new in {\sc idr5} we report ages and distances from the recent literature.

\begin{table*}
\begin{center}
\caption{Open Cluster parameters}
\tiny
\begin{tabular}{llrrrrcccl}
\hline \hline
  \multicolumn{1}{l}{Id} &
  \multicolumn{1}{l}{R.A.} &
  \multicolumn{1}{l}{Dec.} &
  \multicolumn{1}{c}{Age } &
  \multicolumn{1}{c}{R$_{\rm GC}$} &
  \multicolumn{1}{c}{Z } &
  \multicolumn{1}{c}{rv } &
   \multicolumn{1}{c}{[Fe/H]} &
  \multicolumn{1}{l}{n. m.} &
  \multicolumn{1}{l}{Ref. Age \& Distance} \\
   \multicolumn{1}{c}{} &
  \multicolumn{2}{c}{(J2000.0)} &
  \multicolumn{1}{c}{(Gyr)} &
  \multicolumn{1}{c}{(kpc)} &
  \multicolumn{1}{c}{(pc)} &
  \multicolumn{1}{c}{(km s$^{-1}$)} &
  \multicolumn{1}{c}{(dex)} &
 \multicolumn{1}{l}{} &
   \multicolumn{1}{l}{} \\
\hline
NGC6067                               &16:13:11               &     -54:13:06  &   0.10$\pm$0.05  & 6.81$\pm$0.12    & -55$\pm$17 &-39.0$\pm$0.8  & +0.20$\pm$0.08 &  12     &\citet{alonso17}\\
NGC2516                               &07:58:04               &       -60:45:12 &  0.11$\pm$0.01   & 7.98$\pm$0.01    & -97$\pm$4        &+23.6$\pm$1.0  & -0.06$\pm$0.05 & 13 &  \citet{randich17}\\
NGC6259                               &17:00:45               &       -44:39:18 &  0.21$\pm$0.03  & 7.03$\pm$0.01    & -27$\pm$13 &+33.3$\pm$0.9  & +0.21$\pm$0.04 & 11 &  \citet{merm01, dias02}\\
NGC3532                               &11:05:39               &       -58:45:12 &  0.30$\pm$0.10  & 7.85$\pm$0.01    &+11$\pm$4 &+4.9$\pm$0.9  & -0.06$\pm$0.14 & 3 &  \citet{clem11}\\
NGC6705                               &18:51:05               &     -06:16:12  &   0.30$\pm$0.05   & 6.33$\pm$0.16    & -95$\pm$10      &+34.8$\pm$0.7  & +0.16$\pm$0.04 &  16     &\citet{cantat14}\\
NGC6633                               &18:27:15               &      +06:30:30  & 0.52$\pm$0.10   & 7.71$\pm$0.01    &  +52$\pm$2      &  -28.9$\pm$0.9 & -0.01$\pm$0.11 & 10       &\citet{randich17}\\  
NGC4815                               &12:57:59               &     -64:57:36  &   0.57$\pm$0.07   & 6.94$\pm$0.04    & -95$\pm$6        & -29.7$\pm$0.5        & +0.11$\pm$0.01 &  3       &\citet{friel14}\\                                 
Tr23                                        &16:00:50                &      -53:31:23 &   0.80$\pm$0.10  & 6.25$\pm$0.15     & -18$\pm$2        &  -61.3$\pm$0.9  &+0.21$\pm$0.04 & 11         &\citet{jacobson16}\\  
Mel71                               & 07:37:30               &     -12:04:00  &   0.83$\pm$0.18  & 10.50$\pm$0.10     & 210$\pm$20 &+50.8$\pm$1.3  & -0.09$\pm$0.03 &  5     &\citet{salaris04}\\
Be81                                       &19:01:36                &      -00:31:00 &   0.86$\pm$0.10  & 5.49$\pm$0.10     & -126$\pm$7      &  +48.2$\pm$0.5       & +0.22$\pm$0.07 & 13        &\citet{m15}\\
NGC6802                              &19:30:35               &      +20:15:42  & 1.00$\pm$0.10   & 6.96$\pm$0.07    &  +36$\pm$3      &  +11.9$\pm$0.9  & +0.10$\pm$0.02 & 8         &\citet{jacobson16}\\
Rup134			           &17:52:43               &       -29:33:00 &  1.00$\pm$0.20  & 4.60$\pm$0.10    & -100$\pm$10     &-40.9$\pm$0.6  & +0.26$\pm$0.06 & 17 &  \citet{carraro06}\\
NGC6005                              &15:55:48                &      -57:26:12 &   1.20$\pm$0.30  & 5.97$\pm$0.34     & -140$\pm$30    &  -23.5$\pm$1.0 & +0.19$\pm$0.02&  11            &\citet{piatti98} \\
Pis18                                      & 13:36:55               &      -62:05:36 &   1.20$\pm$0.40  & 6.85$\pm$0.17      & +12$\pm$2       & -27.5$\pm$0.6    & +0.22$\pm$0.04& 6          &\citet{piatti98}\\
Tr20                                        &12:39:32                &     -60:37:36  &  1.50$\pm$0.15   & 6.86$\pm$0.01       & +136$\pm$4      & -40.1$\pm$1.0        & +0.15$\pm$0.07  & 37 & \citet{donati14GES}\\
Be44                                      &19:17:12                &    +19:33:00  &  1.60$\pm$0.30    & 6.91$\pm$0.12       & +130$\pm$20    & -8.8$\pm$0.5    & +0.27$\pm$0.06  & 7 &\citet{jacobson16}\\
NGC2420			          &11:05:39               &       -58:45:12 &  2.20$\pm$0.30  & 10.76$\pm$0.20    & +765$\pm$50 &+4.9$\pm$0.9  & -0.13$\pm$0.04 & 22 &  \citet{salaris04, sharma06}\\
Be31				 &06:57:36               &       +08:16:00 &  2.50$\pm$0.30  & 15.16$\pm$0.40    & +340$\pm$30 &+57.5$\pm$0.9  & -0.27$\pm$0.06 & 9 &  \citet{cignoni11}\\
Be25                                      & 06:41:16               &   -16:29:12    & 4.00$\pm$0.50     & 17.60$\pm$1.00          & -1900$\pm$200 & +136.0$\pm$0.8 & -0.25$\pm$0.05 &   6  &\citet{carraro05} \\
NGC2243                             & 06:29:34               &    -31:17:00    & 4.00$\pm$1.20     &  10.40$\pm$0.20         & -1200$\pm$100 & +60.2$\pm$0.5 & -0.38$\pm$0.04  & 16 &\citet{BT06}\\
M67					&08:51:18               &       +11:48:00 &  4.30$\pm$0.50  & 9.05$\pm$0.20    & +405$\pm$40 &+34.7$\pm$0.9  & -0.01$\pm$0.04 & 19 &  \citet{salaris04}\\
Be36				&07:16:06               &       -13:06:00 &  7.00$\pm$0.50  & 11.3$\pm$0.20    & -40$\pm$10 &+62.3$\pm$1.6  & -0.16$\pm$0.10 & 7 &  \citet{donati12}\\
\\
\hline \hline
\end{tabular}
\label{tab_clu_par}
\end{center}
\end{table*}

\begin{table*}
\begin{center}
\caption{Abundance in open clusters, expressed in the form [X/H]. }
\tiny
\begin{tabular}{lrrrrrrr}
\hline \hline
  \multicolumn{1}{c}{Id} &
  \multicolumn{1}{c}{[YII/H]} &
  \multicolumn{1}{c}{[ZrII/H]} &
  \multicolumn{1}{c}{[CeII/H]} &
  \multicolumn{1}{c}{[BaII/H]} &
   \multicolumn{1}{c}{[LaII/H]} &
   \multicolumn{1}{c}{[EuII/H]} &
  \\
\hline 
 NGC6067 	& 0.19$\pm$0.05   		& 0.36$\pm$0.07             & 0.13$\pm$0.07 & 0.46$\pm$0.09& 0.20$\pm$0.06 & 0.19$\pm$0.06\\
 NGC2516 	& 0.04$\pm$0.05   		& 0.52$\pm$0.06             & 0.28$\pm$0.10& 0.17$\pm$0.08 & -				&-\\
 NGC6259 	& 0.14$\pm$0.03   		& 0.28$\pm$0.08             & 0.09$\pm$0.12 & 0.08$\pm$0.05 & 0.13$\pm$0.11&0.20$\pm$0.04\\
 NGC3532 	& -0.09$\pm$0.24   		& -             			     & - 				& 0.02$\pm$0.24 & -				&-\\
 NGC6705 	& 0.05$\pm$0.03   		& 0.22$\pm$0.06             & 0.05$\pm$0.09 & 0.20$\pm$0.11& 0.06$\pm$0.06& 0.15$\pm$0.05\\
 NGC6633 	& 0.01$\pm$0.04   		& 0.15$\pm$0.01             & 0.09$\pm$0.02 & 0.12$\pm$0.11 & 0.06$\pm$0.03& -0.03$\pm$0.13\\
 NGC4815 	& 0.08$\pm$0.07   		& 0.23$\pm$0.10             & 0.06$\pm$0.08 & 0.24$\pm$0.10 & 0.09$\pm$0.07&0.05$\pm$0.08 \\
 Tr23 		& 0.04$\pm$0.04 		& 0.22$\pm$0.04             & 0.00$\pm$0.10 & 0.15$\pm$0.13 & -0.01$\pm$0.04&0.16$\pm$0.08 \\
 Mel71 		& -0.09$\pm$0.01  		& 0.07$\pm$0.02             & 0.04$\pm$0.04 & 0.17$\pm$0.05 & -0.01$\pm$0.03&-0.09$\pm$0.05\\
 Be81 		& 0.23$\pm$0.05             	& 0.30$\pm$0.11             & 0.19$\pm$0.12  & 0.01$\pm$0.08 & 0.24$\pm$0.09& 0.21$\pm$0.08 \\
 NGC6802 	& 0.16$\pm$0.02   		& 0.34$\pm$0.06             & 0.17$\pm$0.07 & 0.16$\pm$0.05 & 0.13$\pm$0.06& 0.11$\pm$0.02\\
 Rup134 	& 0.16$\pm$0.02       		& 0.32$\pm$0.08             & 0.15$\pm$0.08 & 0.04$\pm$0.05 & 0.17$\pm$0.05& 0.22$\pm$0.04\\
 NGC6005 	& 0.12$\pm$0.01   		& 0.24$\pm$0.03             & 0.13$\pm$0.01 & 0.08$\pm$0.06 & 0.03$\pm$0.04&0.12$\pm$0.04 \\
 Pis18 		& 0.06$\pm$0.10     		& 0.25$\pm$0.08             & 0.05$\pm$0.03 & 0.14$\pm$0.08 & 0.09$\pm$0.10& 0.16$\pm$0.02\\
 Tr20 		& 0.13$\pm$0.03 	      & 0.25$\pm$0.05              & 0.13$\pm$0.09 & 0.12$\pm$0.07 & 0.10$\pm$0.05& 0.13$\pm$0.07\\
 Be44 		& 0.34$\pm$0.05             &  -                                     & - 	   			& 0.05$\pm$0.18 & 0.23$\pm$0.07& 0.18$\pm$0.07 \\
 NGC2420 	& -0.14$\pm$0.04  	      & 0.01$\pm$0.06              & 0.04$\pm$0.05 & 0.04$\pm$0.09 & -0.01$\pm$0.04& -0.06$\pm$0.08\\
 Be31 		& -0.32$\pm$0.05            & -0.11$\pm$0.08             & -0.09$\pm$0.09 & -0.32$\pm$0.02& -0.14$\pm$0.07&-0.16$\pm$0.04\\
 Be25 		& -0.24$\pm$0.13            &  0.17$\pm$0.21             & 0.01$\pm$0.22 & -0.28$\pm$0.23 & 0.03$\pm$0.11&0.00$\pm$0.09\\
 NGC2243 	& -0.41$\pm$0.08  	      & -0.21$\pm$0.06            & -0.34$\pm$0.07 &  -  			& -0.34$\pm$0.08	&-0.31$\pm$0.10\\
 M67 		& -0.07$\pm$0.02            & 0.08$\pm$0.08               & 0.01$\pm$0.03 & 0.00$\pm$0.09 & 0.16$\pm$0.19&-0.10$\pm$0.09\\
 Be36 		& -0.18$\pm$0.04            & -0.10$\pm$0.04             & -0.11$\pm$0.05 &  -  			& -0.10$\pm$0.20	&-0.01$\pm$0.07\\
  \hline \hline
\end{tabular}
\end{center}
\label{tab_clu_abu_h}
\end{table*}

\begin{table*}
\begin{center}
\caption{Abundance ratios in open  clusters, expressed in the form [X/Fe]. }
\tiny
\begin{tabular}{lrrrrrrr}
\hline \hline
  \multicolumn{1}{c}{Id} &
  \multicolumn{1}{c}{[YII/Fe]} &
  \multicolumn{1}{c}{[ZrII/Fe]} &
  \multicolumn{1}{c}{[CeII/Fe]} &
  \multicolumn{1}{c}{[BaII/Fe]} &
   \multicolumn{1}{c}{[LaII/Fe]} &
   \multicolumn{1}{c}{[EuII/Fe]} 
  \\
\hline 
  NGC6067 	& 0.14$\pm$0.09 & 0.20$\pm$0.10 & 0.07$\pm$0.10 & 0.39$\pm$0.11 & 0.13$\pm$0.09 & 0.16$\pm$0.10\\
  NGC2516 	& 0.11$\pm$0.06 & 0.58$\pm$0.07 & 0.38$\pm$0.10 & 0.20$\pm$0.08 & -&-\\
  NGC6259 	& -0.01$\pm$0.05 & 0.13$\pm$0.08 & -0.06$\pm$0.12 & -0.04$\pm$0.06 & -0.01$\pm$0.11&-0.05$\pm$0.06\\
  NGC3532   &  0.06$\pm$0.13   & -             & - & 0.17$\pm$0.11 & -&-\\
  NGC6705 	& 0.03$\pm$0.06 & 0.15$\pm$0.08 & -0.01$\pm$0.10 & 0.15$\pm$0.12 & 0.03$\pm$0.08&0.01$\pm$0.08\\
   NGC6633 	& 0.11$\pm$0.10 & 0.24$\pm$0.09 & 0.18$\pm$0.09 & 0.24$\pm$0.14 & 0.15$\pm$0.10&-0.05$\pm$0.16\\
  NGC4815 	& 0.11$\pm$0.12 & 0.29$\pm$0.14 & 0.13$\pm$0.12 & 0.31$\pm$0.13 & 0.12$\pm$0.12&-0.05$\pm$0.13\\
  Tr23 	       & 0.01$\pm$0.05 & 0.22$\pm$0.05 & -0.01$\pm$0.11 & 0.05$\pm$0.14 & -0.03$\pm$0.05&0.00$\pm$0.09\\
  Mel71 	       & 0.08$\pm$0.02 & 0.23$\pm$0.02 & 0.21$\pm$0.04 & 0.34$\pm$0.05 & 0.16$\pm$0.03&-0.05$\pm$0.06\\
  Be81 		& 0.10$\pm$0.08 & 0.19$\pm$0.12 & 0.05$\pm$0.14 & -0.10$\pm$0.10 & 0.11$\pm$0.10&0.00$\pm$0.10\\
  NGC6802 	& 0.15$\pm$0.06 & 0.36$\pm$0.08 & 0.18$\pm$0.09 & 0.16$\pm$0.07 & 0.15$\pm$0.08&-0.01$\pm$0.06\\
  Rup134 	& -0.04$\pm$0.06 & 0.14$\pm$0.09 & -0.02$\pm$0.10 & -0.15$\pm$0.08 & -0.06$\pm$0.08&-0.08$\pm$0.07\\
  NGC6005 	& -0.01$\pm$0.01 & 0.12$\pm$0.03 & 0.03$\pm$0.01 & -0.03$\pm$0.05 & -0.07$\pm$0.04&-0.14$\pm$0.04\\
  Pis18 	       & 0.06$\pm$0.10 & 0.20$\pm$0.07 & 0.04$\pm$0.03 & 0.08$\pm$0.08 & 0.06$\pm$0.10&0.00$\pm$0.02\\
  Tr20 	      & 0.08$\pm$0.07 & 0.20$\pm$0.08 & 0.10$\pm$0.11 & 0.08$\pm$0.09 & 0.04$\pm$0.08&-0.03$\pm$0.09\\
  Be44 		& 0.13$\pm$0.07 & -                           & 0.06$\pm$0.10 & - & 0.05$\pm$0.09 &-0.12$\pm$0.09\\
  NGC2420 	& 0.07$\pm$0.05 & 0.22$\pm$0.07 & 0.21$\pm$0.06 & 0.23$\pm$0.09 & 0.18$\pm$0.05 & 0.00$\pm$0.08\\
  Be31 		& 0.05$\pm$0.05 & 0.28$\pm$0.09 & 0.31$\pm$0.10 & 0.06$\pm$0.04 & 0.23$\pm$0.08 & 0.09$\pm$0.06\\
  Be25 		& 0.11$\pm$0.22 & 0.30$\pm$0.27 & 0.34$\pm$0.28 & 0.05$\pm$0.29 & 0.21$\pm$0.20 & 0.19$\pm$0.19\\
  NGC2243 	& 0.07$\pm$0.09 & 0.28$\pm$0.08 & 0.16$\pm$0.09 & -& 0.19$\pm$0.10 & 0.12$\pm$0.12\\
  M67 		& -0.01$\pm$0.04 & 0.14$\pm$0.08 & 0.05$\pm$0.05 & 0.08$\pm$0.10 & 0.24$\pm$0.19 &-0.19$\pm$0.10\\
  Be36 		& 0.05$\pm$0.20 & 0.18$\pm$0.20 & 0.10$\pm$0.20 & - & 0.11$\pm$0.27 & 0.13$\pm$0.20\\
 \hline \hline
\end{tabular}
\end{center}
\label{tab_clu_abu_fe}
\end{table*}

In Tables~3 and 4, we present the median abundances [X/H] and abundance ratios, [X/Fe]=[log(X/H)-log(X/H)$_{\odot}$)-(log(FeI/H)-log(FeI/H)$_{\odot}$)].  
The latter may slightly differ from the simple subtraction of the median [X/H] and [Fe/H], because the mean metallicity [Fe/H] of each cluster is computed from the global metallicity (FEH column of {\sc idr5}), while   
we have computed [X/Fe] using the iron abundance derived from neutral iron lines (FE1 column).  
In principle, for elements with singly ionised atoms it would be more appropriate to use FeII abundances to obtain their abundance ratios, [X/Fe]. 
We have checked the consistency between log(FeI/H) and log(FeII/H) in our sample, finding an excellent agreement, with a small systematic offset  log(FeII/H)-log(FeI/H)$\sim$0.05~dex which might systematically lower our results by this small amount. However, since the FeII abundances are affected, on average, by larger errors due to the smaller number of FeII lines than of FeI lines, we adopted the FeI abundances to compute the [X/Fe] abundance ratios. The choice of FeI to compute [X/Fe] does not affect the trends or relative comparisons within the Gaia-ESO samples.

\section{The abundances of the s-process  elements}
Following the  literature and, in particular,  \cite{overbeek16} we divide the neutron capture elements into s-elements, those dominated by s-process 
if at least 70\% of their abundance in the Sun is produced by the s-process, and, viceversa, in r-elements, those dominated by the r-process for more than 70\% in the Sun.    
Elements whose nucleosynthesis is not dominated by either  s- or r-process are defined as mixed elements.
We follow the review of  \citet{sneden08}  for what concerns the isotopic composition, the mixture of s- and r-processes  in the Sun and their relative  percentages, but we compare also with the recent redetermination of the solar composition by \citet{bisterzo14}.    
 The isotopic composition and (within parenthesis) the total percentage in the Sun from \citet{sneden08} 
 are reported, for each element,  in Table~\ref{solar_s}. In the last column,  we report also the total percentage of each element in the Sun from \citet{bisterzo14}.  

\begin{table*}
\begin{center}
\caption{Solar percentages and isotopic composition for the s-process elements from \citet{sneden08} and \citet{bisterzo14}. }
\tiny
\begin{tabular}{llll}
\hline \hline
\multicolumn{1}{c}{Element} &
\multicolumn{1}{c}{Isotopes} &
\multicolumn{1}{c}{Sneden (2008)} &
\multicolumn{1}{c}{Bisterzo et al. (2014)}\\
\hline 
Y & $^{89}$Y & 72\% & 72\% \\
Zr & $^{90}$Zr, $^{91}$Zr, $^{92}$Zr, $^{94}$Zr, $^{96}$Zr & 52\%,11\%, 15\%, 17\%, 3\% (81\%) & 66\% \\ 
La & $^{139}$La &  75\% & 76\% \\
Ce &  $^{140}$Ce,  $^{142}$Ce & 90\%, 10\% (81\%) & 84\% \\
Ba &$^{134}$Ba,  $^{135}$Ba, $^{136}$Ba, $^{137}$Ba, $^{138}$Ba  & 3\%, 7\%, 9\%, 12\%, 69\% (85\%) & 85\%\\
\hline \hline
\end{tabular}
\end{center}
\label{solar_s}
\end{table*}

\subsection{The effect of the evolutionary stage: comparing abundances in dwarf and giant stars among cluster member stars}

In a purely differential analysis with respect to the Sun of stars in open clusters, 
\citet{dorazi09} showed that [Ba/Fe] ratios are systematically higher in giant stars than in dwarf stars in a sample of clusters
where both kind of stars have been analysed. This effect is related to the use of a dwarf star, the Sun, as a reference in the analysis of giant stars.

\begin{figure*}
   \centering
  \includegraphics[width=.95\textwidth]{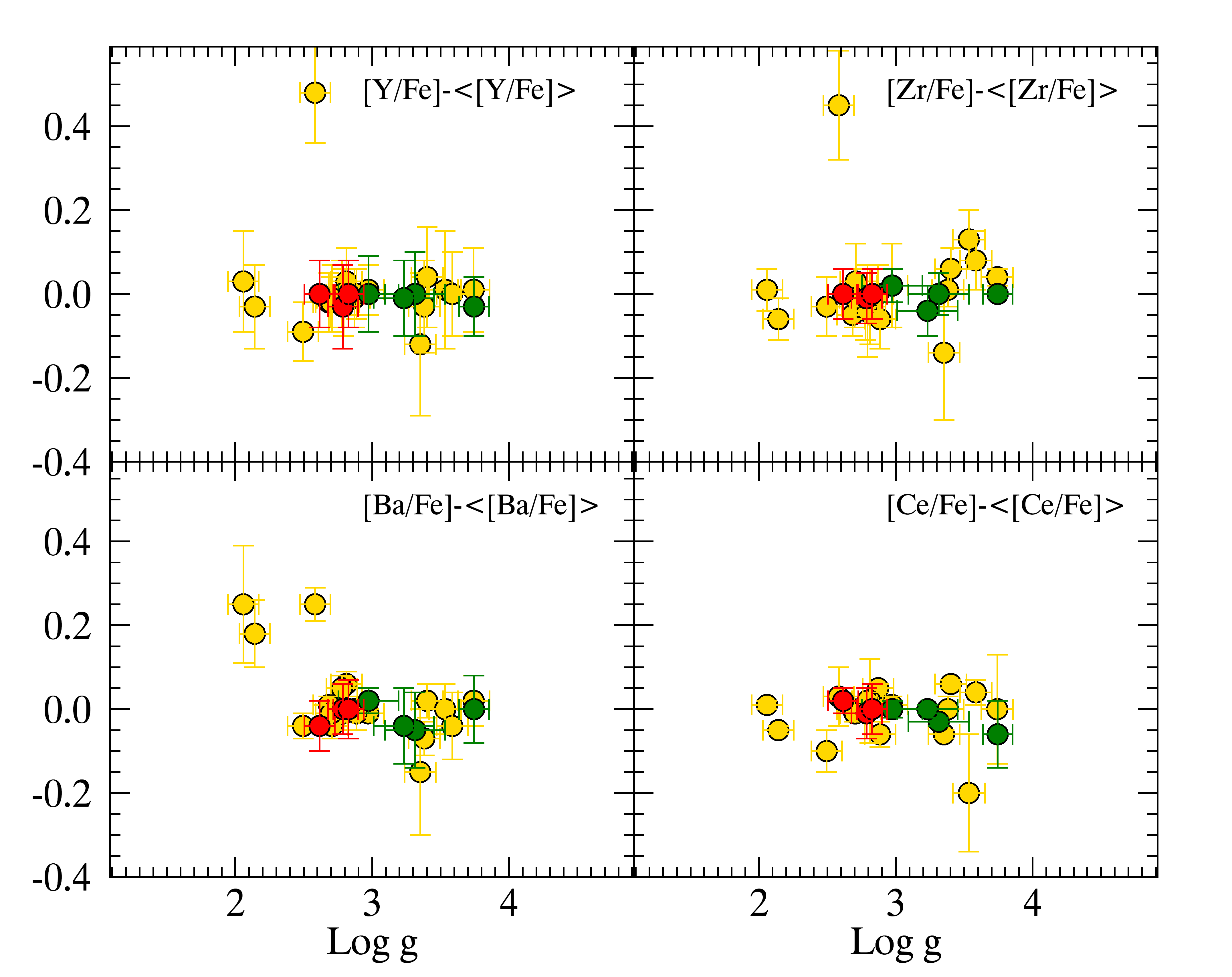}
  \caption{Abundance ratios [X/Fe]  normalised to the median abundance of the cluster  versus $\log$g  in member stars of the open clusters NGC2420 (yellow), NGC6633 (red),  and M67 (green).   Stars with errors in the abundances larger than 0.2~dex are not included in the plot.}
        \label{loggall}
\end{figure*}

\begin{figure}
   \centering
  \includegraphics[width=.52\textwidth]{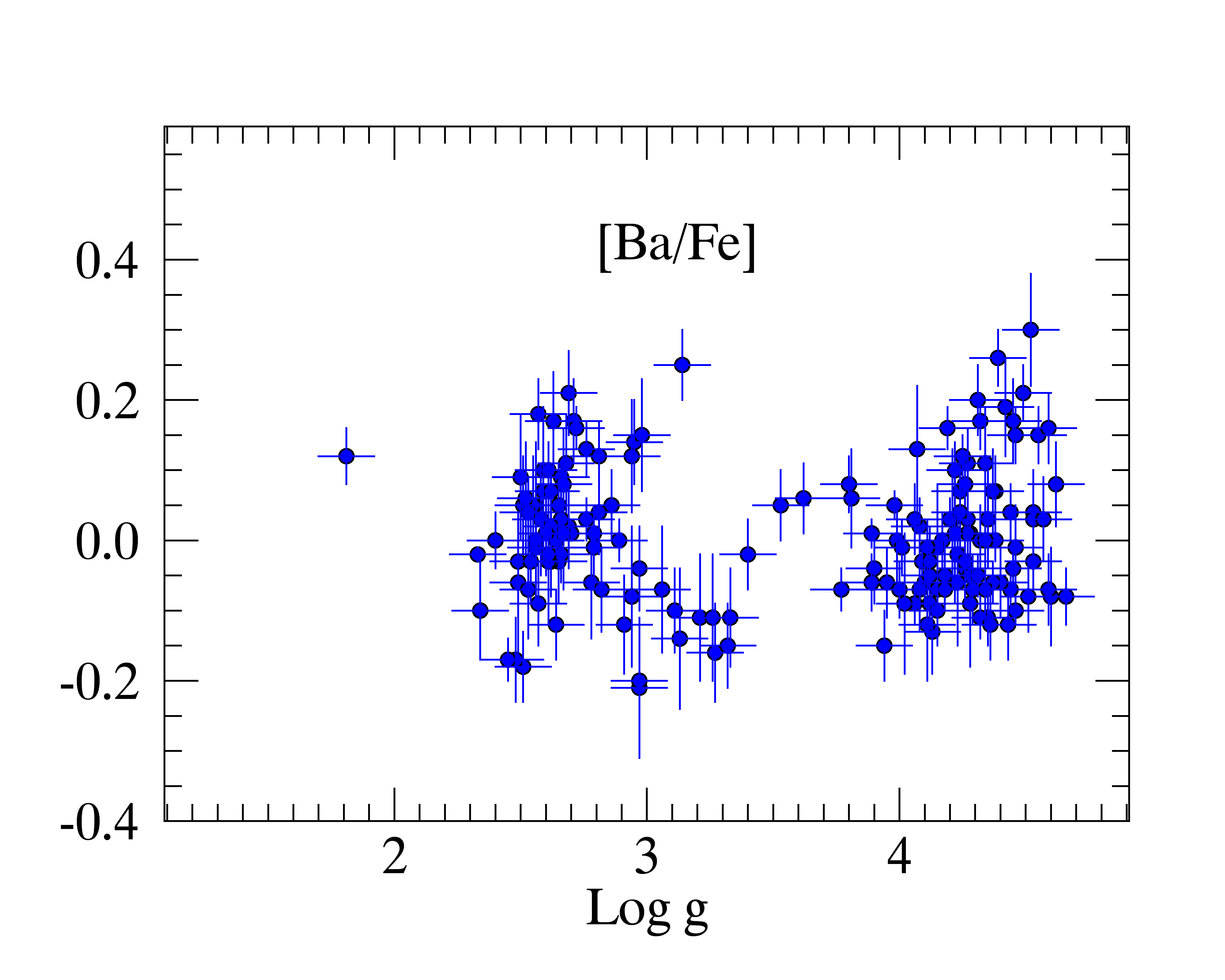}
  \caption{Abundance ratios [Ba/Fe]  versus $\log$g  in Milky Way field stars with -0.1$<$[Fe/H]$<$0.1. }
        \label{loggmw}
\end{figure}

We have analysed the abundance ratios [X/Fe] as a function of the surface gravity in member stars of clusters in  which both dwarf and giant stars were observed. 
The results of [Y/Fe], [Zr/Fe], [Ba/Fe] and [Ce/Fe] for NGC6633, NGC2420,  and M67 are shown in Figure~\ref{loggall}. 
The behaviour of lanthanum is not shown since {\sc idr5} contains La abundances for almost exclusively giant stars.  
 The aim of Figure~\ref{loggall} is to highlight possible differences in the abundances of the s-process elements due the different  stellar parameters.  In these plots we have normalized the abundance ratios of the member stars of each clusters to its median abundance and  we have excluded stars with abundance uncertainties larger than 0.2~dex. 
From this figure, we do not observe, within the uncertainties, any systematic trends of the abundance ratio [Y/Fe], [Zr/Fe], [Ce/Fe] with the surface gravity  for giant and dwarf member stars 
of the same cluster.  
  However, for [Ba/Fe] we notice that the stars with very low gravity in NGC2420 
have Ba abundance 0.2~dex higher than the other member stars. 
In the range of $\log$g from 2.5 to 4 there is no trend of [Ba/Fe] versus $\log$g and thus no 
differences between dwarf and giant stars. 
 To probe how much the overestimation of Ba abundance in low gravity stars might be important, we have selected a sample of Milky Way field stars with -0.1$<$[Fe/H]$<$0.1. 
 In Figure~\ref{loggmw} we plot their [Ba/Fe] versus $\log$g, 
 which indicates that there are no systematic differences between giant and dwarf stars. 
 In addition,  In our sample the number of stars with very low gravity (stars with $\log$g$<$2.5 are $\sim$3\% of the sample of Milky Way field stars) is very small and their eventual over-abundances of Ba should not influence our results.


As a side result, it is worth noticing the presence of a peculiar star in NGC2420:  one of the member stars of NGC2420 with CNAME=07382696+2133313, has systematically higher abundances of Y, Zr, and Ba. Its radial velocity is perfectly consistent with those of the other cluster members. However, in the HR diagram 
it is located slightly outside the main locus of the Red Clump stars and also the abundance of some other elements are somewhat 
discrepant with respect to the average abundance of the cluster (for instance [O/H], [C/H], [Si/H], [Sc/H], [Ti/H] are about -0.3~dex, -0.1~dex, -0.1~dex, -0.15~dex, -0.15~dex respectively,  lower than the average cluster values).  Since its [Fe/H] and radial velocity are consistent with those of the cluster, we can make several hypothesis on the nature of this star:  {\em i)} it is not a cluster member, but a field star with exactly the same metallicity and radial velocity of the cluster; {\em ii)} it is a remnant of binary system in which one of two companions was merged thus modifying the surface composition of the other companion; {\em iii)} the different composition might be due to planet engulfment \citep[see, e.g,][]{spina15}. 
However, the study of this peculiar star is beyond the scope of the present paper and it deserves further investigations.

\subsection{The origin of the s-process elements}

Most of the neutron capture elements are produced in a mixture of astrophysical environments, involving both the r- and s-processes.
There are a few cases in which one of the two processes dominates above the other,  being then the major process responsible for  the element production, but in general
both processes contribute, with one or the other dominating at different epochs and thus in different metallicity ranges. 

\begin{figure*}
   \centering
  \includegraphics[width=.95\textwidth]{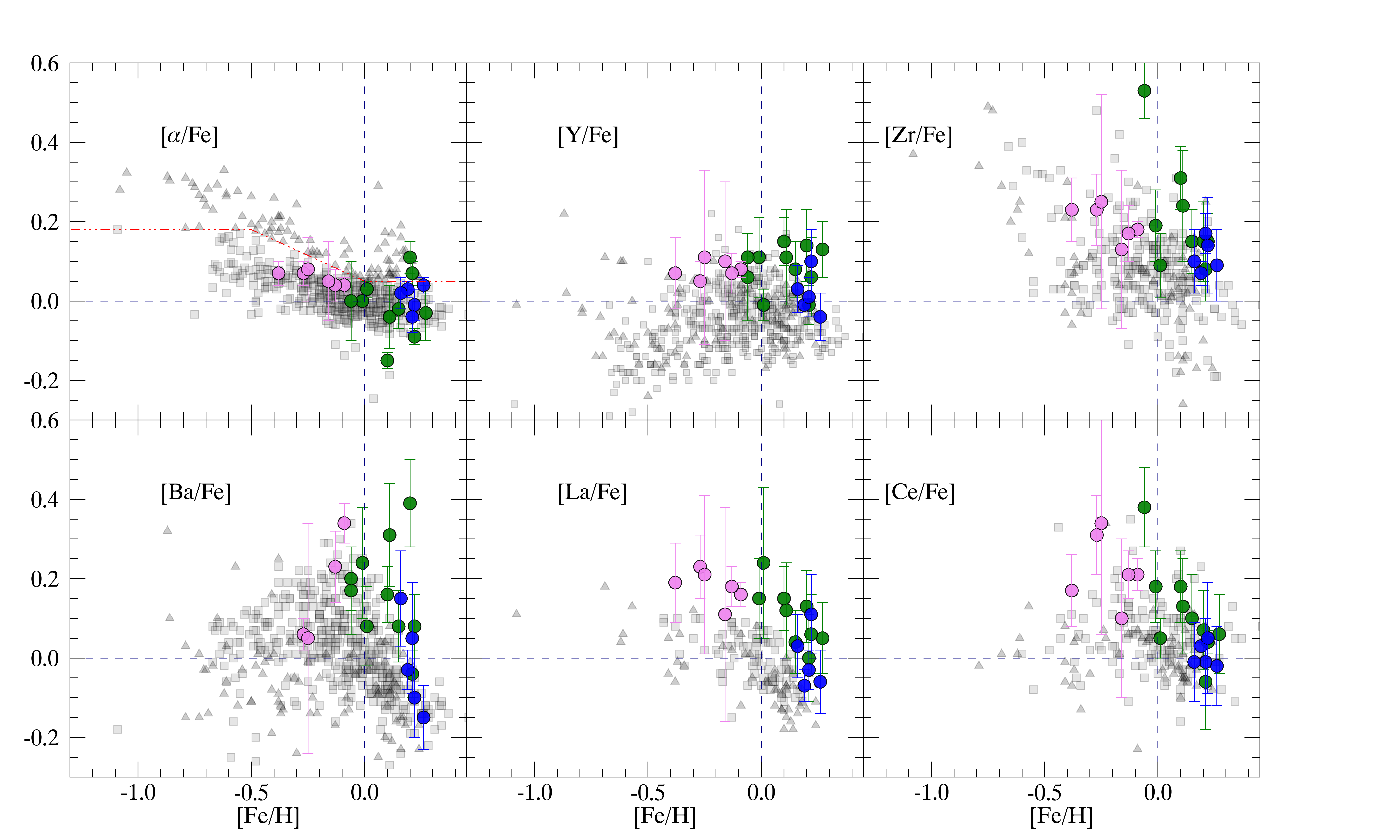}
  \caption{Abundance ratios [X/Fe] versus [Fe/H] in member stars of clusters (large filled circles-- in blue for clusters with R$_{\rm GC}<$6.5 kpc, in green for clusters with 6.5~kpc$<$R$_{\rm GC}<$9.5 kpc, and in violet 
  for clusters with R$_{\rm GC}>$9.5 kpc), in thin disc stars (small grey squares) and 
  in thick disc stars (small black triangles). The blue dashed lines represent the solar values. The red dash-dotted line is the relationship of \citet{adibekyan11} which  separates thin and thick disc stars.} 
        \label{elfeA}
\end{figure*}

\begin{figure*}
   \centering
  \includegraphics[width=.95\textwidth]{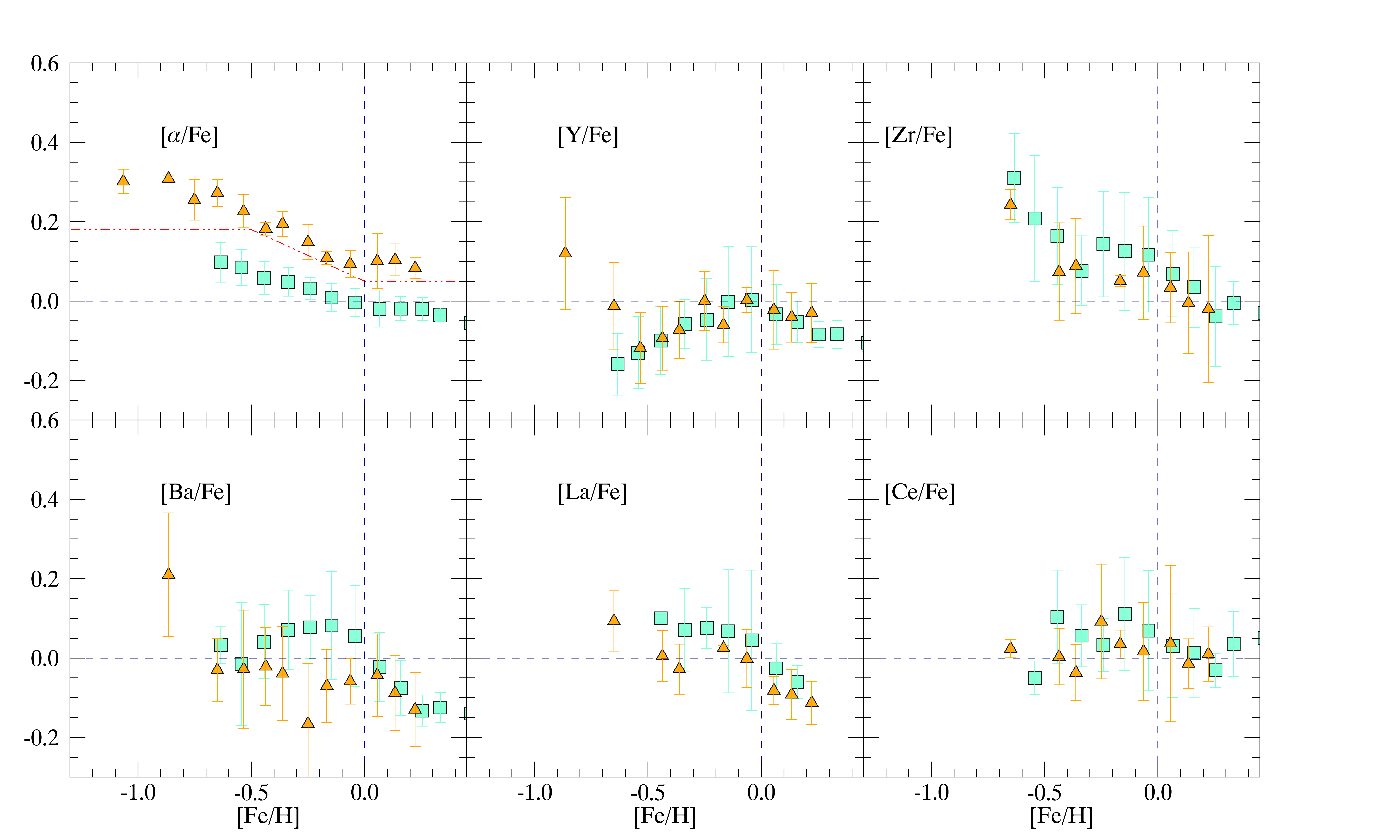}
  \caption{Abundance ratios [X/Fe] versus [Fe/H] in thin disc stars (cyan squares, binned results, bin width 0.1~dex) and 
  in thick disc stars (orange triangles, binned results, bin width 0.1~dex). 
  The blue dashed lines represent the solar values.  The red dash-dotted line is the relationship of \citet{adibekyan11} which  separates thin and thick disc stars.} 
        \label{elfeB}
\end{figure*}

In Figures~\ref{elfeA} and \ref{elfeB},  we show the abundance ratios [X/Fe] of the s-process elements ordered by increasing atomic number plotted with respect to [Fe/H].  
Figure~\ref{elfeA} presents the thin and thick individual stars abundances and the open cluster median abundances: this figure allows us to estimate the differences between field stars and open clusters.  
We note that the age and distance distributions of the cluster and field star samples plotted here differ in important ways.  The majority of field stars have ages between 3 and 7 Gyr, with very few stars younger than 2 Gyr.  By contrast, all but six of the open clusters are younger than 2 Gyr.  In addition, the fields stars have been limited to 1.5 kpc around the solar radius, from 6.5~kpc$<$R$_{\rm GC}<$9.5~kpc, while the open clusters range from 4.5 to 18 kpc in Galactocentric radius.   Consequently, any age and distance dependent trends in abundances will affect the cluster and field star distributions in these diagrams differently.  To help understand the impact of the distance differences, we have color-coded the open clusters by their Galactocentric distance, separating the clusters into radial ranges of R$_{\rm GC}<$6.5~kpc (blue), 6.5~kpc$<$R$_{\rm GC}<$9.5~kpc (green), and R$_{\rm GC}>$9.5~kpc (violet).

Figure~\ref{elfeB} shows the thin and thick abundances, 
averaged in bins 0.1~dex wide: this figure aims at comparing the different Galactic populations and to seek differences between thick and thin disc stars. 
The binned results are shown in the Figures only if there is more than one star in the considered interval. The errors on the binned values are 1-$sigma$ standard deviation of the average.
Typical errors on the individual abundances are: 
0.15~dex for $\log$(Y~{\sc II}/H), 0.15~dex for $\log$(Zr~{\sc II}/H),  0.08~dex for $\log$(Ba~{\sc II}/H), 0.09~dex for $\log$(La~{\sc II}/H), 0.14~dex for $\log$(Ce~{\sc II}/H), and 0.16~dex for $\log$(Eu~{\sc II}/H).  


In the first panel of both Figures~\ref{elfeA} and \ref{elfeB}, we present the [$\alpha$/Fe] vs. [Fe/H] distribution of field stars, with 
the chemical dividing line between thin and thick disc stars. To separate thin and thick stars we have used the relationship of \citet{adibekyan11}. As expected from 
their  young ages, most open clusters belong to the thin disc populations. The separation between the thin and thick disc stars is clear and in agreement 
with previous results  \citep[e.g.][]{rc14}.  

In the second panels, [Y/Fe]  is almost flat across the whole metallicity range, with a relatively small dispersion and a slight decrease towards the lower metallicity. 
The cluster abundances occupy the upper envelope of the region defined by the abundances of field stars, while the distributions of  thin and thick disc stars are coincident (second panel of Figure~\ref{elfeA}).
This difference between cluster and field stars is seen here for the first time with a statistically significant sample of cluster members and of field stars. In the next sections, we show that this is likely an evolutionary effect due to the younger ages of cluster stars with respect to the field population.  

In the third panels,  [Zr/Fe] increases both in thin and thick disc
towards lower metallicity, 
indicating a non-negligible contribution from massive stars, which, given its higher s-process percentage with respect to Y, is a bit unexpected. 
The recent paper by \citet{delgado17} also indicates that Zr shows a clear increasing trend for lower
metallicities. The trend is not new, and it was observed in previous works \citep[e.g.,][]{GH10, mish13, BB16, zhao16}. 
The behaviour of Zr can be related to a reassessment of its s-process contribution. The work of \citet{bisterzo14} indicates that only $\sim$66\% of Zr is produced by the s-process in the Sun. Other processes may contribute to the production of Zr, as discussed by \citet{travaglio04} who introduced the LEPP process associated with massive stars to explain the abundance patterns of the neutron capture elements, especially at low metallicity. 
As in the case of [Y/Fe],  [Zr/Fe] in open clusters is slightly higher with respect to thin and thick  stars, which appear statistically indistinguishable, although both clusters and field stars have [Zr/Fe] elevated above the solar ratio.

With a percentage of 85\%,  Ba (fourth panels of Figures~\ref{elfeA} and \ref{elfeB}) is dominated by the s-process in the metallicity range  from [Fe/H]$\geq$-1.0 to super-solar metallicity. 
Only at lower metallicities, the primary r-process contribution, relatively small at the time of formation of the Sun,  plays an important role \citep{travaglio99}.
In the field stars, the abundance ratio [Ba/Fe] is close to solar for metallicities between [Fe/H]=-1.0 and 0, with a peak of [Ba/Fe]$\sim$0.1~dex in the range -0.4$<$[Fe/H]$<$-0.6.  
In the super-solar metallicity range, we distinguish between [Ba/Fe] in open clusters and field stars; [Ba/Fe]$\geq$0~dex for clusters located in the Solar neighbourhood and 
it is negative for the innermost, more metal rich clusters and for the thin and  thick disc stars. 
The thick disc abundance ratios are systematically lower than the thin disc ratios, with the exception of stars with [Fe/H]$>$0~dex for which thin and thick disc stars have coincident 
[Ba/Fe] distributions. 

Among the second-peak elements, La (fifth panels of Figures~\ref{elfeA} and \ref{elfeB}) has a slight increase of [La/Fe] towards low metallicity. The open clusters 
are located in the upper envelope of the distribution of field star abundance ratios. 
Thin and thick disc stars show similar patterns, with thin disc stars slightly overabundant in [La/Fe].  

Finally, the behaviour of [Ce/Fe] (sixth panels of Figures~\ref{elfeA} and \ref{elfeB}) is almost flat in the [Fe/H] range considered. 
At variance with [Ba/Fe], [Ce/Fe] is constant also at super solar metallicities. Thin and thick stars share similar patterns.
  For both elements, Ce and La, the more metal-poor open clusters have higher [La/Fe] and [Ce/Fe] than the more metal-rich ones.  However, these metal-poor clusters are located in the outer disc and their abundances may be more a reflection of the radial gradient of these elements in the Galactic disc.

\subsection{Estimating the r-process contribution}
\begin{figure*}
   \centering
  \includegraphics[width=1.0\textwidth]{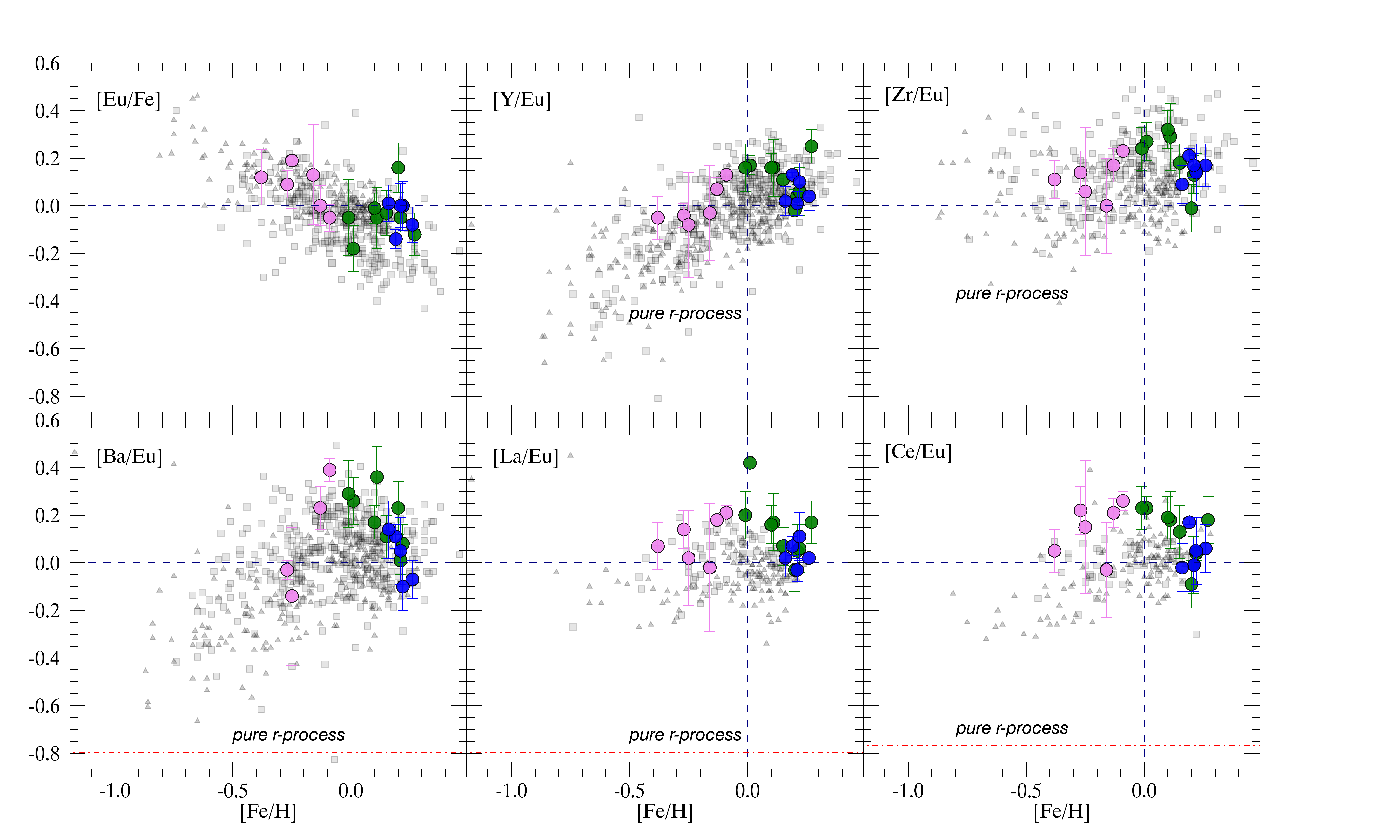}
  \caption{[Eu/Fe] and abundance ratios [X/Eu]  versus [Fe/H]. Symbol and colour codes  as in Fig.~\ref{elfeA}. 
The blue dashed lines represent the solar values, while the red dot-dashed lines indicate the abundance ratio below which  the pure r-contribution dominates, derived using the percentages of \citet{bisterzo14} and the solar abundances of \citet{grevesse07}.    }
 \label{xeu_perA}
 \end{figure*}

\begin{figure*}
   \centering
  \includegraphics[width=1.0\textwidth]{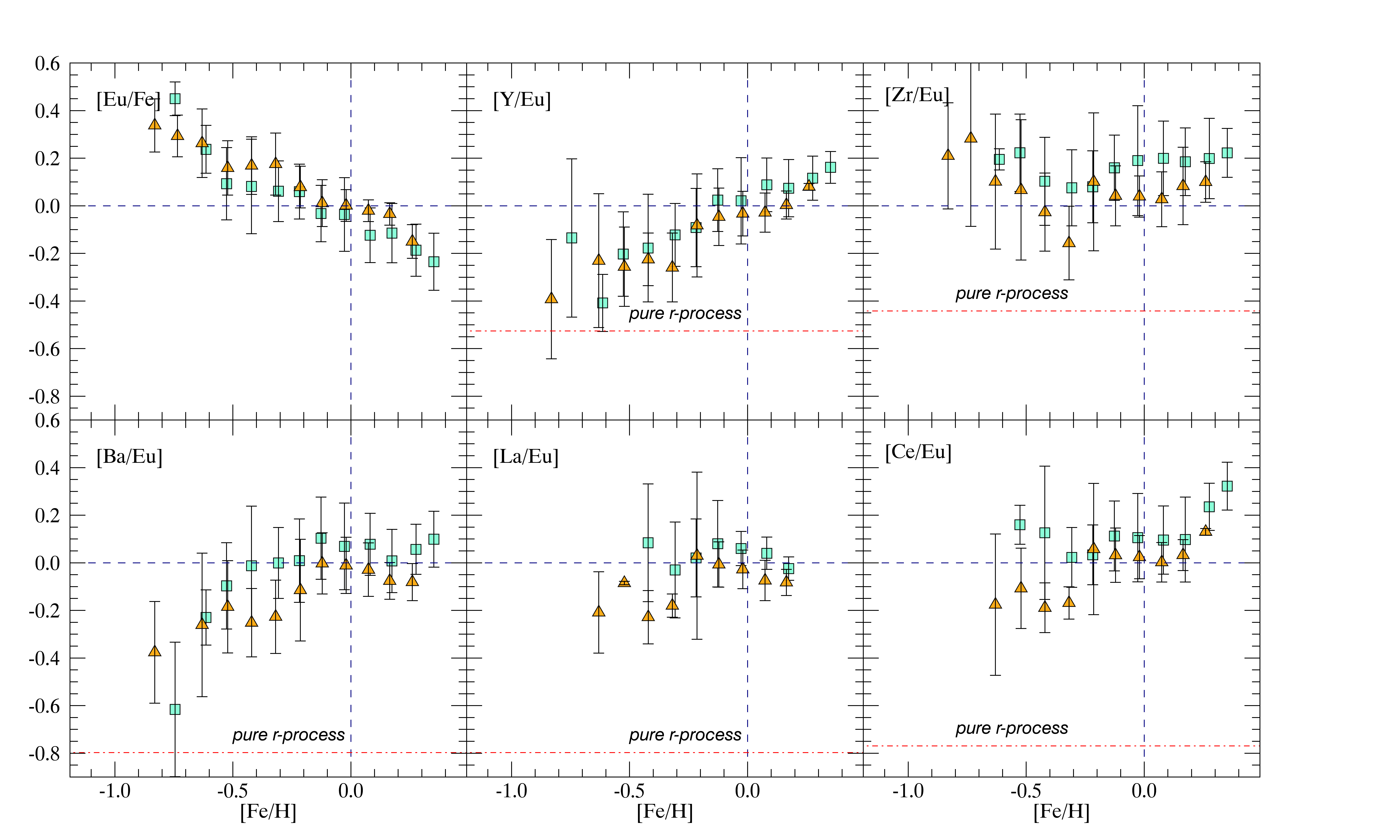}
  \caption{[Eu/Fe] and abundance ratios [X/Eu]  versus [Fe/H]. Symbol and colour codes  as in Fig.~\ref{elfeB}.   The lines are coded as in Figure~\ref{xeu_perA}.   
   }
 \label{xeu_perB}
 \end{figure*}

The abundance ratio [X/Eu] can be used as a direct way to judge the relative importance of the r-process channels during the evolution  of the Galaxy in the metallicity range of our sample stars, as has been done in previous works \citep[e.g., ][]{MG01, BB16}. 
In Figures~\ref{xeu_perA}  and \ref{xeu_perB}   we present the [X/Eu] abundance ratios versus [Fe/H].   
In the plots the blue dashed lines represent the s-process
solar values, while the red dot-dashed line indicates 
the pure r-process contribution of both elements, derived using the percentages of \citet{bisterzo14} and the solar abundances of \citet{grevesse07}.
The two figures highlight different aspects: in the former we have a direct comparison between field stars and open clusters, and in the latter the comparison between thin and thick disc populations. 

In the first panels of both figures, we present the abundance ratio [Eu/Fe] versus [Fe/H]. Europium is an r-process element and it shows a clear increase towards the lower metallicity. 
Although the first hypothesis about the mechanisms responsible of r-process nucleosynthesis date back many years ago \citep{burbidge57, seeger65}, 
there is still an open debate about the dominant production site of the r-process elements \citep[see e.g.][]{thielemann10}.
The r-process nucleosynthesis must be associated with both the evolution of massive stars ($M>$10 M$_{\odot}$) and  
with neutron star mergers \citep{matteucci14}. 
Very recently, observations of the optical counterpart of a gravitational wave source 
have identified the presence of r-process elements in the spectrum of the merger of two neutron stars, clearly pointing to the production of these elements in neutron star merger events \citep{cote17}.

In the second panels, for the first-peak element  Y  the sample shows an increasing contribution from s-process elements above a pure r-process level, but there are some stars in the range -1.0$<$[Fe/H]$<$-0.5 that appear to have a pure r-process composition. These stars belong to both the thin and thick discs. Over the full metallicity range, the abundance ratios of the thick and thin discs are offset from each other by about 0.1~dex in [Y/Eu]. 

In the third panels, we present the abundance ratio [Zr/Eu] versus [Fe/H], which is almost constant, quite far from the pure r-process contribution, and slightly above the solar value. Almost all the open clusters have positive [Zr/Eu]. For this element, the thin and thick disc populations are separated: the abundance ratios of thick disc stars are lower, although  they follow a similar pattern to the thin disc stars (see third panel of Figure~\ref{xeu_perB}).
In the fourth panels, barium shows a mild increase of the r-process component towards 
low metallicity. In the thin disc [Ba/Eu] is close to solar values from [Fe/H]$\sim$+0.4 down to [Fe/H]$\sim$-0.5~dex, then it decreases. For the thick disc, the decline starts at higher metallicity, [Fe/H]$\sim$-0.2~dex, and again, over the entire metallicity range, the thick disk stars have lower abundance ratio. [Ba/Eu] in open clusters is positive in the range -0.2$<$[Fe/H]$<$+0.2~dex, while it is close to zero for the clusters more metal poor and more metal rich. 

In the fifth panels, [La/Eu] is quite flat over the whole metallicity range. The open clusters are all slightly over solar in [La/Eu], and thick disk abundance ratios are slightly lower than in thin disk stars.  

In the sixth panels, [Ce/Eu] is also quite flat over the whole metallicity range and it 
tends to increase in both thin and thick disc stars towards high [Fe/H]. 
The few stars at low metallicity [Fe/H]$\leq$-0.5~dex might point to a decrease of [Ce/Eu], indicating the beginning of the r-process contribution in the thick disc. 

We can thus conclude that in the considered  metallicity range, the production of the five elements is  clearly dominated by the s-process.  At any metallicity, the thick disk stars show very slightly lower [s/Eu] ratios than the thin disk, consistent with their slightly higher [Eu/Fe] values.

 \section{The time evolution of the abundances of the s-process elements}
 \label{sec:time}
 
One of the most challenging issues in the field of chemical evolution is the nature of the time-evolution of the s-process abundances. 
The classical chemical and chemo-dynamical models \citep{PT97,travaglio99, travaglio04, raiteri99}, based on the accepted scenario for s-processing at that time, predicted 
a plateau or even a decrease in the abundances [X/Fe] versus age after the  solar  formation.  A late increase was not expected. 
However, recently, after the original work of \citet{dorazi09}, a number of works have claimed the presence of a continuous increase of
the first- and second-peak element abundances, sometimes with a further growth starting 5-6 Gyr ago \citep{maiorca11, dasilva12, nissen16, spina16, nissen17, silva17, feltzing17}.

In the present work, we combine the open cluster sample, which basically maps the recent epochs in the evolution of the Galactic disc, with the field stars, divided into thin and thick disc on the basis of their [$\alpha$/Fe], as explained in Section~\ref{sec_stars}.  
The ages of the field stars, computed with the isochrone projection methods,  are much less accurate than the ages of open clusters, and they have only a statistical meaning. 
 However, keeping in mind the limit of measuring stellar ages, the Gaia-ESO {\sc idr5} sample gives us the opportunity, for the first time, to combine the largest sample of s-process element abundances 
 measured in a completely homogeneous way in very young and very old stars. 
 No biases due to, for instance, the line list selection, the method of deriving abundances, or adopted reference abundances are present in the Gaia-ESO {\sc idr5}. 
  
\begin{figure*}
   \centering
  \includegraphics[width=1.0\textwidth]{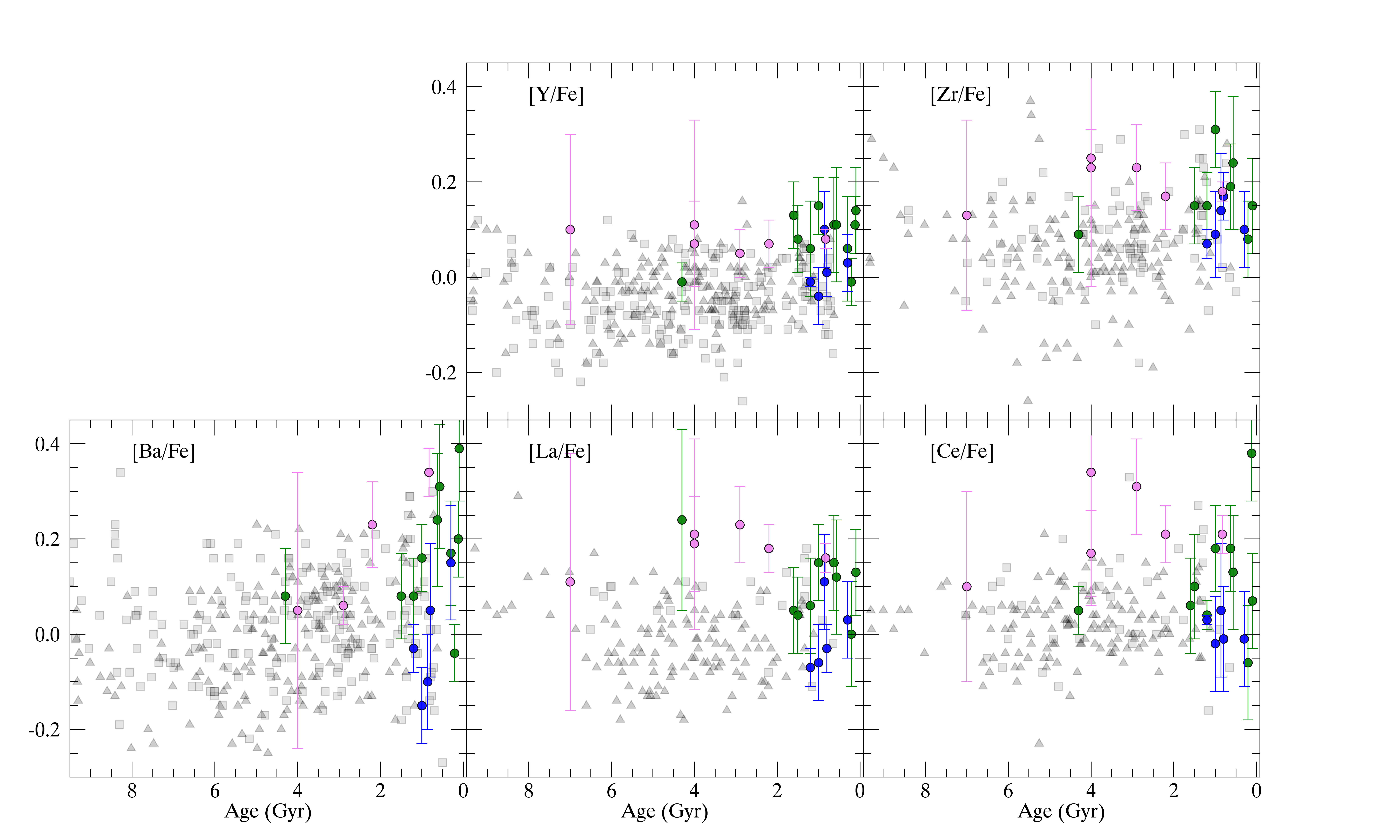}
  \caption{Abundance ratios [X/Fe]  versus Age. Symbol and colour codes as in Fig.~\ref{elfeA}.   }
 \label{elfe_ageA}
 \end{figure*}

\begin{figure*}
   \centering
  \includegraphics[width=1.0\textwidth]{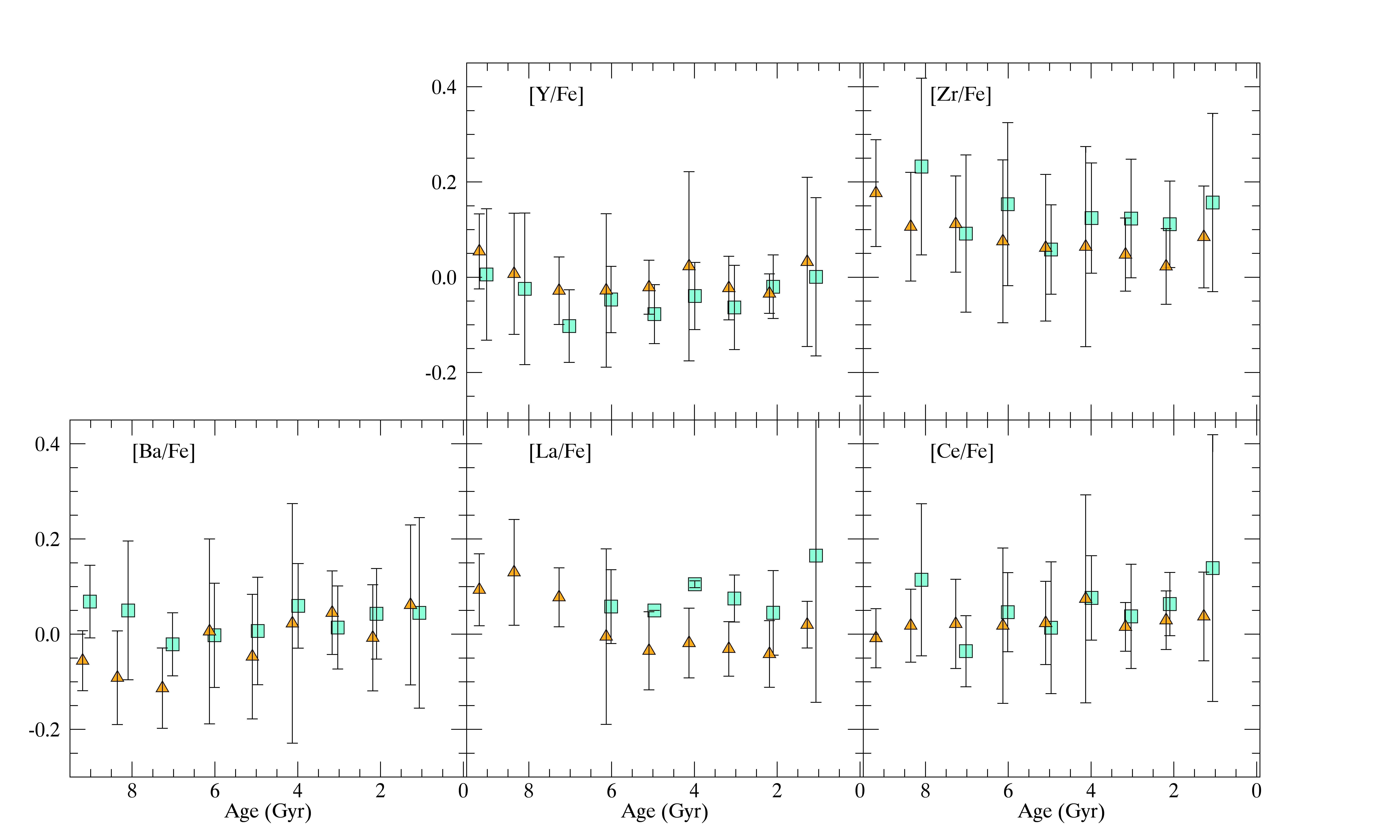}
  \caption{Abundance ratios [X/Fe]  versus Age. Symbol and colour codes as in Fig.~\ref{elfeB}.    }
 \label{elfe_ageB}
 \end{figure*}

\begin{figure*}
   \centering
  \includegraphics[width=1.0\textwidth]{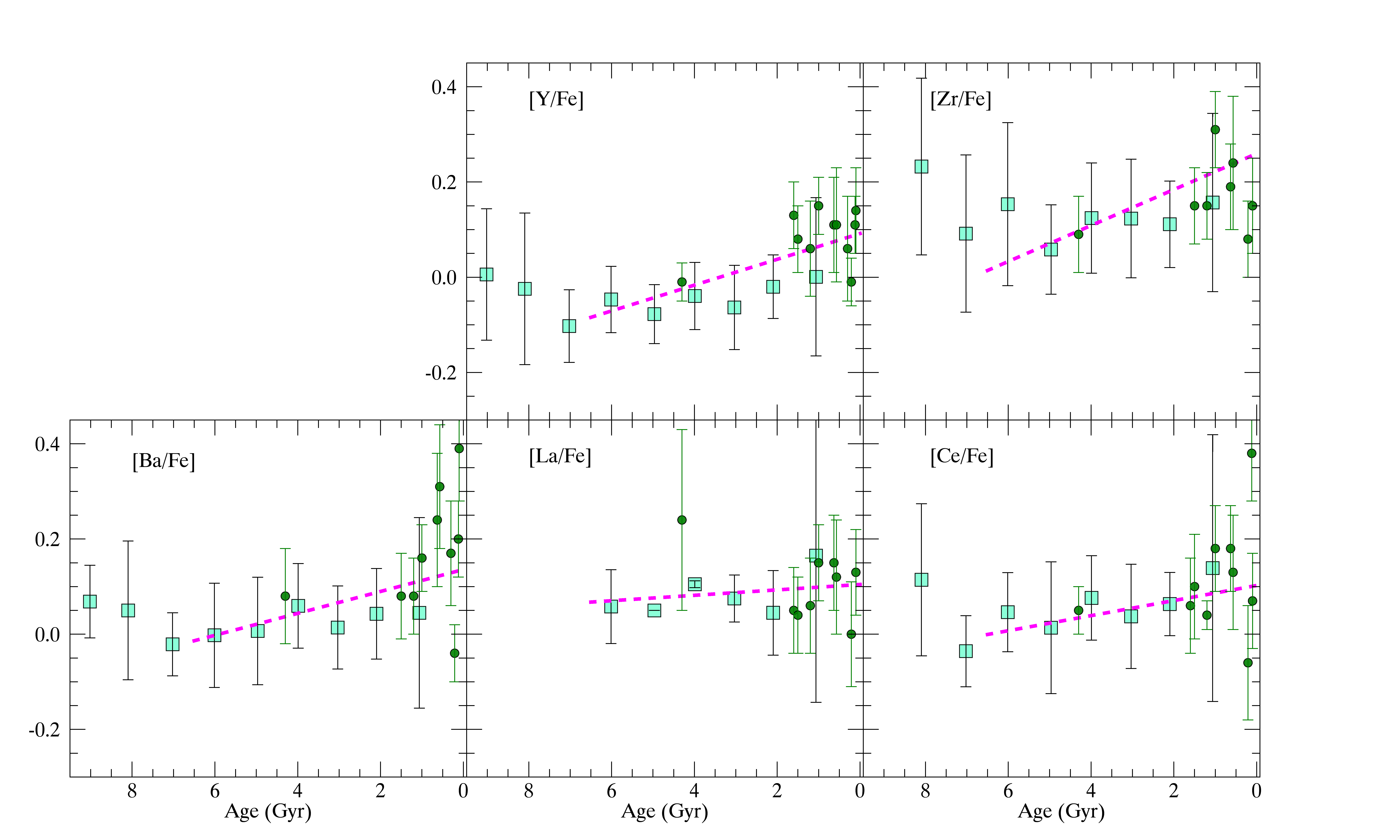}
  \caption{Abundance ratios [X/Fe]  versus Age for the thin disc stars (binned results, bin 0.1~dex wide, cyan squares) and the open clusters located in the solar neighborhood (green circles).  The magenta dashed lines are the weighted linear fits to the cumulative sample of 
  solar neighbourhood clusters and thin disc field stars (see Table~\ref{tab:fits} for the coefficients) for stellar ages $<$8~Gyr.   }
 \label{elfe_ageC}
 \end{figure*}

In Figure~\ref{elfe_ageA}, ~\ref{elfe_ageB}, and ~\ref{elfe_ageC} we show the abundance ratios [X/Fe]  versus age of both field stars and open clusters.  Recall that the sample of thin and thick disk stars is already restricted to the Solar neighbourhood, i.e. 6.5~kpc$<$R$_{\rm GC}<$9.5~kpc. The open clusters span a much larger range of distances, and we have distinguished them by radial bins according to their Galactocentric location, denoting inner disk, solar neighbourhood, and outer disk clusters separately in the figures. 
In Figure~\ref{elfe_ageA} we compare the field stars with the open clusters, while in Figure~\ref{elfe_ageB} we compare the thin and thick disc [X/Fe] versus stellar ages (binning in age bins  1~Gyr wide), and finally in Figure~\ref{elfe_ageC}
we limit our analysis to the Solar neighbourhood thin disc populations, 
and open clusters 
belonging to the same radial range. 

Figure~\ref{elfe_ageA} aims at highlighting the behaviour of clusters located at different Galactocentric distance with respect to the field stars. 
The Solar neighbourhood and inner disc clusters populate the youngest age regions, while the outer disc clusters span a large range of ages. 
For all s-process elements, we notice a difference between the  youngest clusters of the Solar neighbourhood and the inner disc clusters. The former usually have  higher [X/Fe] than the latter. The outer disc open clusters typically have  [X/Fe] similar to the Solar neighbourhood clusters.  
Dividing the cluster sample by Galactocentric distance makes clear that there is a strong dependence of the heavy s-element abundances on 
the location of the clusters. 
The behaviour of [X/Fe] with Galactocentric radius seems to be complex and likely related to different star formation histories and to metallicity dependency of the stellar yields.

Figure~\ref{elfe_ageB} shows [X/Fe] as a function of stellar ages for the binned thin and thick disc populations.  In the radial range shown here, i.e. 6.5~kpc~$<$R$_{\rm GC}<$9.5~kpc, there are not strong differences 
between the behaviours of the thin and thick disc populations. All elements seem to indicate a slight increase in [X/Fe] starting at ages $<$8~Gyr. These trends are most clear in the thin disk, where the sample sizes per bin are larger and mean values are perhaps better determined than for the thick disc.  

Figure~\ref{elfe_ageC} presents [X/Fe] versus stellar age for the thin disc and cluster populations in the Solar neighbourhood with weighted linear fits to the combined cluster and thin disc samples for ages lower than 8~Gyr, i.e. corresponding to epoch in which we expect the contribution of the s-process starts. 
In Table~\ref{tab:fits} we present these weighted linear fits, reporting the slopes, the intercepts, and correlation coefficients for fits by element. 
Restricting our sample to the Solar neighborhood allows us to better appreciate the increasing trends with decreasing stellar ages.  
The first peak elements, [Y/Fe] and [Zr/Fe], have increases of 0.15 to 0.25 dex over the past 6-7 Gyr.  These slopes agree with those of \citet{spina18}, who find slopes of $-0.029$ and $-0.027$ dex~Gyr{$^-1$} for Y and Zr, respectively. 
The elements of the second peak also show a slight increase with age, more pronounced for [Ba/Fe] and [Ce/Fe], and flatter for  [La/Fe] (see Table~\ref{tab:fits}). Slopes for Ba, at $-0.03$ dex~Gyr{$^-1$}, agree with those found by \citet{dasilva12,spina18}. In the case of La, which is measured only in giant stars, we have a much smaller sample of stars and thus the measurement of the trend is less well constrained (see its relative error  on the slope and its correlation coefficient in Table~\ref{tab:fits}).  

The growth in the s-process abundances in the youngest Galactic stellar populations has been reproduced  by, for instance, the chemical evolution model of \citet{maiorca12}, which was one of the first models that made an attempt to explain it. 
Their approach was to relate the s-process enhancement to a strong contribution to the production of these elements from low-mass stars, which start to contribute later in the lifetime of the Galaxy. 
In particular, they maintain that the observed enhancements might be produced during the nucleosynthesis processes in the asymptotic giant branch (AGB) phases of low-mass stars ($M<$1.5~M$_{\odot}$) 
under the hypothesis that they release neutrons from the $^{13}$C($\alpha$,n)$^{16}$O reaction to create larger reservoirs  of neutrons than in more massive AGB stars ($M>$1.5~M$_{\odot}$). 
Later studies have given  a physical explanation for the necessity of having larger reservoirs  of neutrons, the so-called $^{13}$C pocket, in low-mass stars. 
The extended $^{13}$C pocket requires the existence of very efficient mixing episodes and the transport mechanisms most commonly adopted such as 
overshoot \citep{herwig00, cristallo09}, gravity waves \citep{denissenkov03, battino16}, rotationally driven shear or thermohaline mixing 
do not suffice \citep{palmerini11}, but magnetic buoyancy can explain the process \citep{busso07, nordhaus08, trippella16, TLC17}. 

The observed slope of [Ba/Fe] as a function of stellar age is somewhat steeper than those of [La/Fe] and [Ce/Fe], although their intercepts indicate similar maximum global enrichment ($\sim$ 0.1 dex) for all elements, in agreement with their similar nucleosynthesis. 
However the maximum abundances of [Ba/Fe] reached in the youngest open clusters is higher than [La/Fe] and [Ce/Fe]. A larger enrichment of Ba compared to La is difficult
to reconcile considering similar contributions to both elements by s- and r-processes. 
\citet{mish15} discussed several possibilities to explain [Ba/La] in young open clusters, which can reach $\sim$0.15~dex. They proposed
an additional contribution from an intermediate neutron capture process, the so-called i-process. The i-process is characterised by neutron densities in the order
of 10$^{15}$ neutrons cm$^{-3}$ and is triggered by the mixing or ingestion of H in He-burning stellar layers \citep{cr77, bertolli13}. 
The i-process can potentially explain this larger ratio and 
the higher [Ba/La] in the youngest objects might suggest that 
the i-process contribution is becoming more relevant
during the most recent few Gyr.

 

\begin{table}
\begin{center}
\tiny
\caption{Coefficients and Pearson indexes of the weighted linear fits for [X/Fe] versus stellar age (age$<$8~Gyr).  }
\begin{tabular}{llll}
\hline\hline
Element  & Slope (dex~Gyr$^{-1}$)    & Intercept (dex)             &  R \\   
\hline
Yttrium [Y/Fe]	& -0.023$\pm$0.009  & 0.13$\pm$0.03  & -0.6 \\	
Zirconium [Zr/Fe]	& -0.038$\pm$0.013  & 0.26$\pm$0.03  & -0.5 \\
Barium  [Ba/Fe] & -0.027$\pm$0.007  & 0.09$\pm$0.02  & -0.8 \\
Lanthanum [La/Fe]	& -0.005$\pm$0.015  & 0.10$\pm$0.04  & -0.2 \\
Cerium [Ce/Fe]	& -0.016$\pm$0.010  & 0.10$\pm$0.03  & -0.5 \\
\hline \hline
\end{tabular}
\end{center}
\label{tab:fits}
\end{table}

 \subsection{The s-process relative indicator [Ba/Y]}
 
\begin{figure}
   \centering
  \includegraphics[width=.50\textwidth]{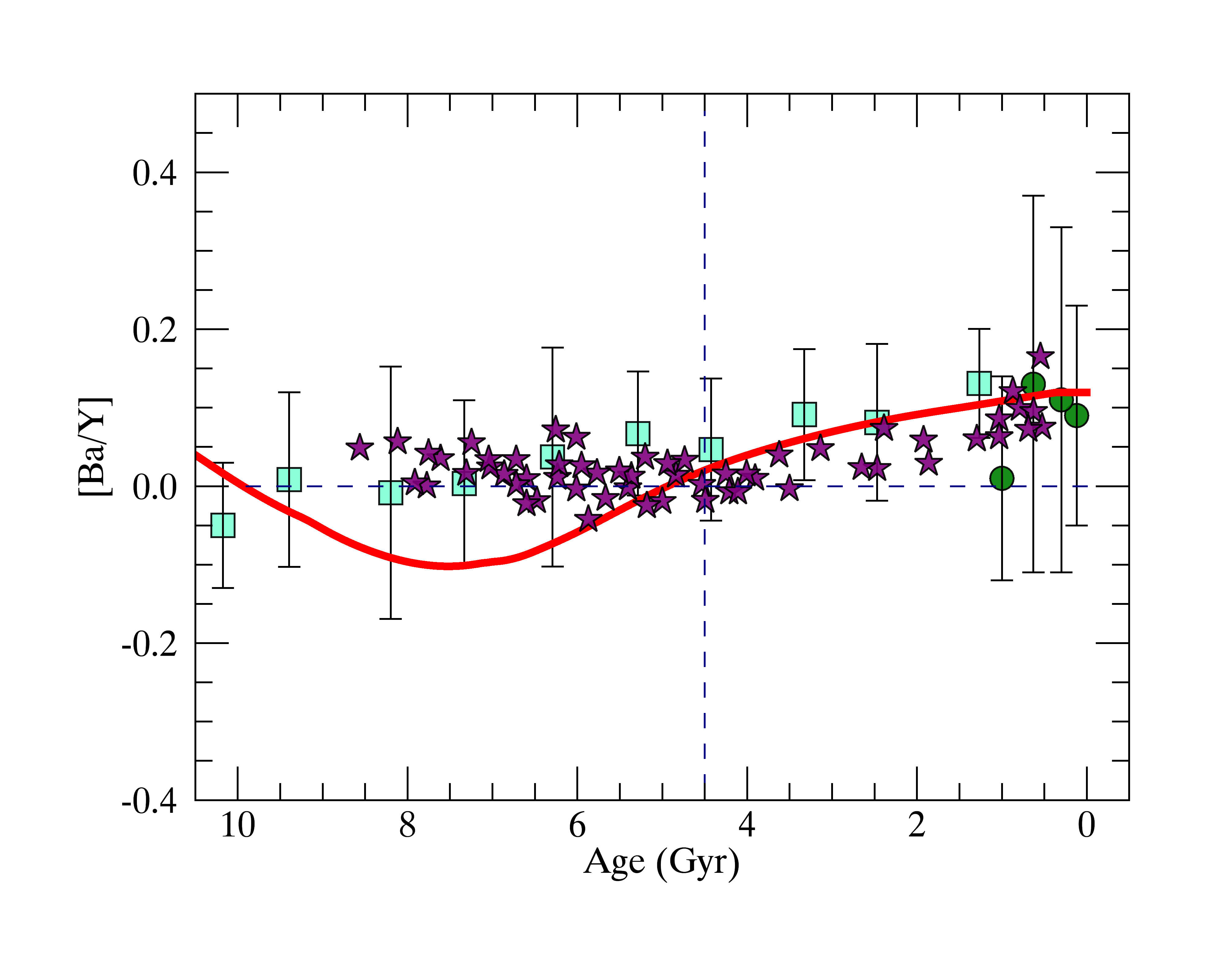}
  \caption{Abundance ratios [Ba/Y]  versus Age in cluster (green circles) and thin disc stars (cyan squares, bins of 1 Gyr) in the radial range 6.8~kpc$<$R$_{\rm GC} <$~8.2~kpc and -0.1$<$[Fe/H]$<$~+0.1, and in the solar twin stars of \citet{spina18} (purple stars).  In red the model curve of \citet{maiorca12}.  }
 \label{bay_age}
 \end{figure}

In Figure~\ref{bay_age} we show [Ba/Y] versus stellar ages. 
 Y and Ba belong to the first and second peak, respectively, and thus they allow us to estimate the relative weight of the production of light and heavy s-process elements with time.
  We have chosen these two elements because they are measured in many more stars than Zr, La, and Ce. 
  In the plot we compare our results with the sample of solar twin stars of \citet{spina18}. 
 To do that, we have selected thin disc stars and open clusters with the same characteristics of the 
  \citet{spina18}'s sample: 6.8~kpc$<$R$_{\rm GC} <$~8.2~kpc and -0.1$<$[Fe/H]$<$~+0.1. 
The two samples are in very good agreement, both indicate for [Ba/Y]  an increase from about $\sim$6~Gyr to the youngest ages. The oldest stars show a slight growth of [Ba/Y] together with an increasing dispersion. 
In this Figure, we have also plotted the results of the model of \citet{maiorca12} for the time evolution of [Y/Ba]. The model results are  in qualitative agreement, whilst it predicts a lower [Ba/Y] for ages larger of $\sim$6~Gyr. As already noticed by \citet{spina18}, the behaviour of [Ba/Y] versus stellar age is directly related to dependence of the AGB star yields on metallicity. 
The interplay during the chemical evolution of the Milky Way between very low-mass AGB stars, with enhanced reservoir of $^{13}$C, and more massive AGB stars with the usual 
 $^{13}$C pocket, might produce the observed behaviour of [Ba/Y] (see, e.g., Figure 3 of \citet{maiorca12}). 
 The observational result can be explained also by the AGB models of \citet{KL16}, which predict that the [hs/ls] ratio in the winds of a typical
 AGB star is -0.026 at solar metallicity and +0.320 at  half solar metallicity. 
 
\begin{figure*}
   \centering
  \includegraphics[width=1.00\textwidth]{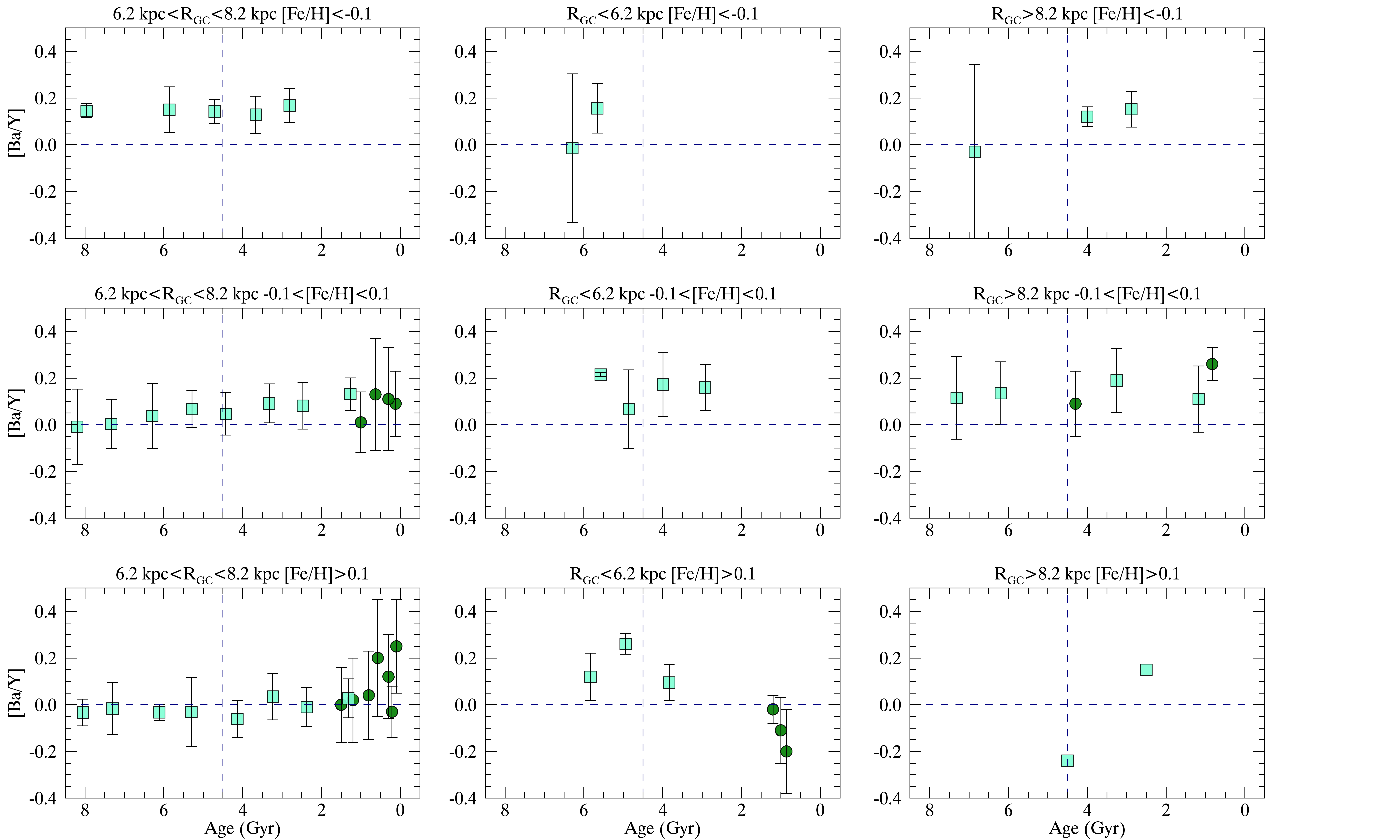}
  \caption{Abundance ratios [Ba/Y]  versus Age in cluster and thin disc stars (symbols as in Figure~\ref{bay_age} ) the three radial ranges:  6.2~kpc$<$R$_{\rm GC} <$~8.2~kpc (left panels),  R$_{\rm GC} <$~6.2~kpc (central panels) and 
  R$_{\rm GC} >$~8.2~kpc (right panels). For each radial bin, we divide the plot in three metallicity ranges: below solar [Fe/H]$<$~-0.1 (panels on the upper row), solar -0.1$<$[Fe/H]$<$~+0.1 (panels on the central row) and super solar [Fe/H]$>$~+0.1 (panels on the lower row). }  
 \label{bay_age_all}
 \end{figure*}

In Figure~\ref{bay_age_all}, we expand the study of the behaviour of [Ba/Y] versus stellar ages to different radial and metallicity bins.
We basically divide the sample in three radial ranges: the solar neighbourhood  6.2~kpc$<$R$_{\rm GC} <$~8.2~kpc,  the inner disc  R$_{\rm GC} <$~6.2~kpc and 
the outer disc  R$_{\rm GC} >$~8.2~kpc. Each radial range is then divided in metallicity bins:  below solar [Fe/H]$<$~-0.1, solar -0.1$<$[Fe/H]$<$~+0.1 and super solar [Fe/H]$>$~+0.1. 
The relative weight of the s-process indicator [Ba/Y] clearly depends on the Galactocentric position, and thus on the infall time scales and star formation history, and on the metallicity. 
The growth of [Ba/Y] with stellar ages is present in the solar neighbourhood, but it is less clear and less constrained in the inner disc and in outer disc. 
The metallicity dependence is very strong, with [Ba/Y] usually higher in the [Fe/H]$<$~-0.1 range, and decreasing towards the higher metallicities. 
All this information is new, and it was not available at the time of previous studies. It provides important constraints to the  study of the s-process production.

\section{Summary and conclusions}
\label{summary}
Taking advantage of the large and homogeneous sample of stellar parameters and elemental abundances of Gaia-ESO {\sc idr5}, we study the abundances of five s-process elements in the thin and the thick discs and in the population of open clusters. 
We compute statistical ages for field stars and we perform membership analysis of stars in open clusters, computing the median abundances of each cluster. 
We estimate the effect of evolutionary stages in the s-process abundances by comparing giant and dwarf stars, which are members of the same clusters, and we do not find any remarkable difference. 
We identify the abundance ratio patterns versus [Fe/H] of both field and open cluster stars, and their behaviour versus stellar ages. 
Using our large and homogenous sample of open clusters, thin and thick disc stars, spanning an age range larger than 10~Gyr, 
and selecting a sample of clusters and field stars in the solar ring, 
we confirm an increase towards young ages of  the five analysed s-process abundances.
The trend is more clear for [Y/Fe], [Zr/Fe] and [Ba/Fe], while it has a lower correlation coefficient for [La/Fe] and [Ce/Fe] perhaps because of the limited number of stars in which these elements are measured.
Clusters located in the inner and outer disc have different abundance ratios with respect to those of the solar neighbourhood. 

The global growth of the s-process abundance ratio with time can be explained by a strong contribution to the production of these elements from low-mass stars, which start to contribute later in the lifetime of the Galaxy \citep[see, e.g.][]{maiorca12}. 
However, the differences in the maximum values 
reached in open clusters for Ba and La point to the necessity of an additional production mechanism for Ba and they might be explained by the occurrence of the i-process at later epochs in the evolution of the Galaxy \citep[e.g.][]{mish15}. 
The new observations of Gaia-ESO have confirmed with a large statistical sample the behaviour of s-process elements in the Solar neighbourhood, and now they 
give us  the possibility to investigate the interplay among the different neutron-capture processes, the metallicity dependence of their stellar yields, and the physical process involved in the creation of this important group of elements in different parts of our Galaxy.

    \begin{acknowledgements}
    We are grateful to the referee for her/his comments and suggestions that improved the quality of the paper. 
These data products have been processed by the Cambridge Astronomy Survey Unit (CASU) at the Institute of Astronomy, University of Cambridge, and by the FLAMES/UVES reduction team at INAF/Osservatorio Astrofisico di Arcetri. These data have been obtained from the Gaia-ESO Survey Data Archive, prepared and hosted by the Wide Field Astronomy Unit, Institute for Astronomy, University of Edinburgh, which is funded by the UK Science and Technology Facilities Council.
This work was partly supported by the European Union FP7 programme through ERC grant number 320360 and by the Leverhulme Trust through grant RPG-2012-541. We acknowledge the support from INAF and Ministero dell' Istruzione, dell' Universit\`a' e della Ricerca (MIUR) in the form of the grant "Premiale VLT 2012". The results presented here benefit from discussions held during the Gaia-ESO workshops and conferences supported by the ESF (European Science Foundation) through the GREAT Research Network Programme.
F.J.E. acknowledges financial support from ASTERICS project (ID:653477, H2020-EU.1.4.1.1. - Developing new world-class research infrastructures).
S. D. acknowledges support from Comit\'e Mixto ESO-GOBIERNO DE CHILE.
AB thanks for support from the Millenium Science Initiative, Chilean Ministry of Economy.
This project has received funding from the European Union's Horizon 2020 research and innovation programme under the Marie Sklodowska-Curie grant agreement No 664931. 
E.D.M., V.A. and S.G.S. acknowledge the support from Funda\c{c}\~ao para a Ci\^encia e a Tecnologia (FCT, Portugal) through the research grant through national funds and by FEDER through COMPETE2020 by grants UID/FIS/04434/2013 \& POCI-01-0145-FEDER-007672, PTDC/FIS-AST/1526/2014 \& POCI-01-0145-FEDER-016886 and PTDC/FIS-AST/7073/2014 \& POCI-01-0145-FEDER-016880. E.D-M., V.A. and S.G.S also acknowledge support from FCT through Investigador FCT contracts nr.  IF/00849/2015/CP1273/CT0003, IF/00650/2015/CP1273/CT0001 and IF/00028/2014/CP1215/CT0002.
T.B. was supported by the project grant 'The New Milky' from the Knut and Alice Wallenberg foundation.
T. M. acknowledges support provided by the Spanish Ministry
of Economy and Competitiveness (MINECO) under grant AYA-2017-88254-P

\end{acknowledgements}


\begin{thebibliography}{}

\bibitem[Adibekyan et al.(2011)]{adibekyan11} Adibekyan, V.~Z., Santos, N.~C., Sousa, S.~G., \& Israelian, G.\ 2011, \aap, 535, LL11 
\bibitem[Alonso-Santiago et al.(2017)]{alonso17} Alonso-Santiago, J., Negueruela, I., Marco, A., et al.\ 2017, \mnras, 469, 1330 
\bibitem[Asplund et al.(2009)]{agss09} Asplund, M., Grevesse, N., Sauval, A. J. \& Scott, P. 2009, \araa, 47, 481 
\bibitem[Battino et al.(2016)]{battino16} Battino, U., Pignatari, M., Ritter, C., et al.\ 2016, \apj, 827, 30 
\bibitem[Battistini \& Bensby(2016)]{BB16} Battistini, C., \& Bensby, T.\ 2016, \aap, 586, A49 
\bibitem[Bensby et al.(2014)]{bensby14} Bensby, T., Feltzing, S., \& Oey, M.~S.\ 2014, \aap, 562, A71 
\bibitem[Bertolli et al.(2013)]{bertolli13} Bertolli, M.~G., Herwig, F., Pignatari, M., \& Kawano, T.\ 2013, arXiv:1310.4578 
\bibitem[Bisterzo et al.(2014)]{bisterzo14} Bisterzo, S., Travaglio, C., Gallino, R., Wiescher, M., \& K{\"a}ppeler, F.\ 2014, \apj, 787, 10 
\bibitem[Bisterzo et al.(2017)]{bisterzo17} Bisterzo, S., Travaglio, C., Wiescher, M., K{\"a}ppeler, F., \& Gallino, R.\ 2017, \apj, 835, 97 
\bibitem[Bragaglia \& Tosi(2006)]{BT06} Bragaglia, A., \& Tosi, M.\ 2006, \aj, 131, 1544 
\bibitem[Bressan et al.(2012)]{bressan12} Bressan, A., Marigo, P., Girardi, L., et al.\ 2012, \mnras, 427, 127 
\bibitem[Burbidge et al.(1957)]{burbidge57} Burbidge, E.~M., Burbidge, G.~R., Fowler, W.~A., \& Hoyle, F.\ 1957, Reviews of Modern Physics, 29, 547 
\bibitem[Busso et al.(1999)]{busso99} Busso, M., Gallino, R., \& Wasserburg, G.~J.\ 1999, \araa, 37, 239 
\bibitem[Busso et al.(2007)]{busso07} Busso, M., Wasserburg, G.~J., Nollett, K.~M., \& Calandra, A.\ 2007, \apj, 671, 802 
\bibitem[Cantat-Gaudin et  al.(2014)]{cantat14} Cantat-Gaudin, T., Vallenari, A., Zaggia, S., et al.\ 2014, \aap, 569, AA17 
\bibitem[Carraro et al.(2005)]{carraro05} Carraro, G., Geisler, D., Moitinho, A., Baume, G., \& V{\'a}zquez, R.~A.\ 2005, \aap, 442, 917 
\bibitem[Carraro et al.(2006)]{carraro06} Carraro, G., Janes, K.~A., Costa, E., \& M{\'e}ndez, R.~A.\ 2006, \mnras, 368, 1078 
\bibitem[Cignoni et al.(2011)]{cignoni11} Cignoni, M., Beccari, G., Bragaglia, A., \& Tosi, M.\ 2011, \mnras, 416, 1077 
\bibitem[Clem et al.(2011)]{clem11} Clem, J.~L., Landolt, A.~U., Hoard, D.~W., \& Wachter, S.\ 2011, \aj, 141, 115 
\bibitem[C{\^o}t{\'e} et al.(2017)]{cote17} C{\^o}t{\'e}, B., Fryer, C.~L., Belczynski, K., et al.\ 2017, arXiv:1710.05875 
\bibitem[Cowan \& Rose(1977)]{cr77}Cowan J. J., Rose W. K., 1977, ApJ, 212, 149
\bibitem[Cristallo et al.(2009)]{cristallo09} Cristallo, S., Straniero, O., Gallino, R., et al.\ 2009, \apj, 696, 797 
\bibitem[da Silva et al.(2012)]{dasilva12} da Silva, R., Porto de Mello, G.~F., Milone, A.~C., et al.\ 2012, \aap, 542, A84 
\bibitem[Delgado Mena et al.(2017)]{delgado17} Delgado Mena, E., Tsantaki, M., Adibekyan, V.~Z., et al.\ 2017, \aap, 606, A94 
\bibitem[Dell'Omodarme et al.(2012)]{do12} Dell'Omodarme, M., Valle, G., Degl'Innocenti, S., \& Prada Moroni, P.~G.\ 2012, \aap, 540, A26 
\bibitem[Denissenkov et al.(2013)]{denissenkov03} Denissenkov, P.~A., Herwig, F., Truran, J.~W., \& Paxton, B.\ 2013, \apj, 772, 37 
\bibitem[Dias et al.(2002)]{dias02} Dias, W.~S., Alessi, B.~S., Moitinho, A., \& L{\'e}pine, J.~R.~D.\ 2002, \aap, 389, 871 
\bibitem[Donati et al.(2012)]{donati12} Donati, P., Bragaglia, A., Cignoni, M., Cocozza, G., \& Tosi, M.\ 2012, \mnras, 424, 1132 
\bibitem[Donati et al.(2014)]{donati14GES} Donati, P., Cantat Gaudin, T., Bragaglia, A., et al.\ 2014b, \aap, 561, AA94
\bibitem[D'Orazi et al.(2009)]{dorazi09} D'Orazi, V., Magrini, L., Randich, S., et al.\ 2009, \apjl, 693, L31 
\bibitem[Feltzing et al.(2017)]{feltzing17} Feltzing, S., Howes, L.~M., McMillan, P.~J., \& Stonkut{\.e}, E.\ 2017, \mnras, 465, L109 
\bibitem[Friel et al.(2014)]{friel14} Friel, E.~D., Donati, P., Bragaglia, A., et al.\ 2014, \aap, 563, AA117 
\bibitem[Gaia Collaboration et al.(2016)]{gaia} Gaia Collaboration, Prusti, T., de Bruijne, J.~H.~J., et al.\ 2016, \aap, 595, A1 
\bibitem[Gallino et al.(1998)]{gallino98} Gallino, R., Arlandini, C., Busso, M., et al.\ 1998, \apj, 497, 388 
\bibitem[Gilmore et al.(2012)]{gilmore12} Gilmore, G., Randich, S., Asplund, M., et al.\ 2012, The Messenger, 147, 25 
\bibitem[Gonz{\'a}lez Hern{\'a}ndez et al.(2010)]{GH10} Gonz{\'a}lez Hern{\'a}ndez, J.~I., Israelian, G., Santos, N.~C., et al.\ 2010, \apj, 720, 1592 
\bibitem[Grevesse et al.(2007)]{grevesse07} Grevesse, N., Asplund, M., \& Sauval, A.~J.\ 2007, \ssr, 130, 105 
\bibitem[Herwig(2000)]{herwig00} Herwig, F.\ 2000, \aap, 360, 952 
\bibitem[Karakas \& Lugaro(2016)]{KL16} Karakas, A.~I., \& Lugaro, M.\ 2016, \apj, 825, 26 
\bibitem[Kordopatis et al.(2011)]{kor11} Kordopatis, G., Recio-Blanco, A., de Laverny, P., et al.\ 2011, \aap, 535, A107 
\bibitem[Kordopatis et al.(2015a)]{kordo15} Kordopatis, G., Wyse, R.~F.~G., Gilmore, G., et al.\ 2015b, \aap, 582, A122 
\bibitem[Jacobson \& Friel(2013)]{JF13} Jacobson, H.~R., \& Friel, E.~D.\ 2013, \aj, 145, 107 
\bibitem[Jacobson et al.(2011)]{jacobson11} Jacobson, H.~R., Friel, E.~D., \& Pilachowski, C.~A.\ 2011, \aj, 141, 58 
\bibitem[Jacobson et al.(2016)]{jacobson16} Jacobson, H.~R., Friel, E.~D., Jilkova, L., et al.\ 2016, \aap, 591, 37 
\bibitem[Liu et al.(2016)]{liu16} Liu, F., Asplund, M., Yong, D., et al.\ 2016, \mnras, 463, 696 
\bibitem[Maiorca et al.(2011)]{maiorca11} Maiorca, E., Randich, S., Busso, M., Magrini, L., \& Palmerini, S.\ 2011, \apj, 736, 120 
\bibitem[Maiorca et al.(2012)]{maiorca12} Maiorca, E., Magrini, L., Busso, M., et al.\ 2012, \apj, 747, 53 
\bibitem[Magrini et al.(2009)]{magrini09} Magrini, L., Sestito, P., Randich, S., \& Galli, D.\ 2009, \aap, 494, 95 
\bibitem[Magrini et al.(2015)]{m15} Magrini, L., Randich, S., Donati, P., et al.\ 2015, \aap, 580, A85 
\bibitem[Magrini et al.(2017)]{magrini17} Magrini, L., Randich, S., Kordopatis, G., et al.\ 2017, \aap, 603, A2 
\bibitem[Mashonkina \& Gehren(2001)]{MG01} Mashonkina, L., \& Gehren, T.\ 2001, \aap, 376, 232 
\bibitem[Matteucci et al.(2014)]{matteucci14} Matteucci, F., Romano, D., Arcones, A., Korobkin, O., \& Rosswog, S.\ 2014, \mnras, 438, 2177 
\bibitem[Mermilliod et al.(2001)]{merm01} Mermilliod, J.-C., Clari{\'a}, J.~J., Andersen, J., Piatti, A.~E., \& Mayor, M.\ 2001, \aap, 375, 30 
\bibitem[Mishenina et al.(2013)]{mish13} Mishenina, T., Korotin, S., Carraro, G., Kovtyukh, V.~V., \& Yegorova, I.~A.\ 2013, \mnras, 433, 1436 
\bibitem[Mishenina et al.(2015)]{mish15} Mishenina, T., Pignatari, M., Carraro, G., et al.\ 2015, \mnras, 446, 3651 
\bibitem[Modigliani et al.(2004)]{modigliani04} Modigliani, A., Mulas, G., Porceddu, I., et al.\ 2004, The Messenger, 118, 8 
\bibitem[Nissen(2016)]{nissen16} Nissen, P.~E.\ 2016, \aap, 593, A65 
\bibitem[Nissen et al.(2017)]{nissen17} Nissen, P.~E., Silva Aguirre, V., Christensen-Dalsgaard, J., et al.\ 2017, arXiv:1710.03544 
\bibitem[Nordhaus et al.(2008)]{nordhaus08} Nordhaus, J., Busso, M., Wasserburg, G.~J., Blackman, E.~G., \& Palmerini, S.\ 2008, \apjl, 684, L29 
\bibitem[{\"O}nehag et al.(2014)]{on14} {\"O}nehag, A., Gustafsson, B., \& Korn, A.\ 2014, \aap, 562, A102 
\bibitem[Overbeek et al.(2016)]{overbeek16} Overbeek, J.~C., Friel, E.~D., \& Jacobson, H.~R.\ 2016, \apj, 824, 75 
\bibitem[Pagel \& Tautvaisiene(1997)]{PT97} Pagel, B.~E.~J., \& Tautvaisiene, G.\ 1997, \mnras, 288, 108 
\bibitem[Palmerini et al.(2011)]{palmerini11} Palmerini, S., Cristallo, S., Busso, M., et al.\ 2011, \apj, 741, 26 
\bibitem[Pancino et al.(2017)]{pancino17} Pancino, E., Lardo, C., Altavilla, G., et al.\ 2017, \aap, 598, A5 
\bibitem[Pasquini et al.(2002)]{pasquini02} Pasquini, L., Avila, G., Blecha, A., et al.\ 2002, The Messenger, 110, 1 
\bibitem[Pasquini et al.(2008)]{pasquini} Pasquini, L., Biazzo, K., Bonifacio, P., Randich, S., \& Bedin, L.~R.\ 2008, \aap, 489, 677 
\bibitem[Piatti et al.(1998)]{piatti98} Piatti, A.~E., Clari{\'a}, J.~J., Bica, E., Geisler, D., \& Minniti, D.\ 1998, \aj, 116, 801 
\bibitem[Pignatari et al.(2010)]{pignatari10} Pignatari, M., Gallino, R., Heil, M., et al.\ 2010, \apj, 710, 1557 
\bibitem[Raiteri et al.(1993)]{raiteri93} Raiteri, C.~M., Gallino, R., Busso, M., Neuberger, D., \& Kaeppeler, F.\ 1993, \apj, 419, 207 
\bibitem[Raiteri et al.(1999)]{raiteri99} Raiteri, C.~M., Villata, M., Gallino, R., Busso, M., \& Cravanzola, A.\ 1999, \apjl, 518, L91 
\bibitem[Randich et al.(2006)]{randich06} Randich, S., Sestito, P., Primas, F., Pallavicini, R., \& Pasquini, L.\ 2006, \aap, 450, 557 
\bibitem[Randich et al.(2013)]{RG13} Randich, S., Gilmore, G., \& Gaia-ESO Consortium 2013, The Messenger, 154, 47 
\bibitem[Randich et al.(2017)]{randich17} Randich, S., Tognelli, E., Jackson, R., et al.\ 2017, arXiv:1711.07699 
\bibitem[Recio-Blanco et al.(2014)]{rc14} Recio-Blanco, A., de Laverny, P., Kordopatis, G., et al.\ 2014, \aap, 567, A5 
\bibitem[Reddy \& Lambert(2017)]{RL17} Reddy, A.~B.~S., \& Lambert, D.~L.\ 2017, \apj, 845, 151 
\bibitem[Sacco et al.(2014)]{sacco14} Sacco, G.~G., Morbidelli, L., Franciosini, E., et al.\ 2014, \aap, 565, A113 
\bibitem[Salaris et al.(2004)]{salaris04} Salaris, M., Weiss, A., \& Percival, S.~M.\ 2004, \aap, 414, 163 
\bibitem[Schlegel et al.(1998)]{schlegel} Schlegel, D.~J., Finkbeiner, D.~P., \& Davis, M.\ 1998, \apj, 500, 525 
\bibitem[Seeger et al.(1965)]{seeger65} Seeger, P.~A., Fowler, W.~A., \& Clayton, D.~D.\ 1965, \apjs, 11, 121 
\bibitem[Sharma et al.(2006)]{sharma06} Sharma, S., Pandey, A.~K., Ogura, K., et al.\ 2006, \aj, 132, 1669 
\bibitem[Silva Aguirre et al.(2017)]{silva17} Silva Aguirre, V., Lund, M.~N., Antia, H.~M., et al.\ 2017, \apj, 835, 173 
\bibitem[Smiljanic et al.(2014)]{smi14} Smiljanic, R., Korn, A.~J., Bergemann, M., et al.\ 2014, \aap, 570, A122 
\bibitem[Sneden et al.(2008)]{sneden08} Sneden, C., Cowan, J.~J., \& Gallino, R.\ 2008, \araa, 46, 241 
\bibitem[Spina et al.(2015)]{spina15} Spina, L., Palla, F., Randich, S., et al.\ 2015, \aap, 582, L6 
\bibitem[Spina et al.(2016)]{spina16} Spina, L., Mel{\'e}ndez, J., Karakas, A.~I., et al.\ 2016, \aap, 593, A125 
\bibitem[Spina et al.(2018)]{spina18} Spina, L., Mel{\'e}ndez, J., Karakas, A.~I., et al.\ 2018, \mnras, 474, 2580 
\bibitem[Stonkut{\.e} et al.(2016)]{sto16} Stonkut{\.e}, E., Koposov, S.~E., Howes, L.~M., et al.\ 2016, \mnras, 460, 1131 
\bibitem[Thielemann et al.(2011)]{thielemann10} Thielemann, F.-K., Arcones, A., K{\"a}ppeli, R., et al.\ 2011, Progress in Particle and Nuclear Physics, 66, 346 
\bibitem[Tognelli et al.(2011)]{tognelli11} Tognelli, E., Prada Moroni, P.~G., \& Degl'Innocenti, S.\ 2011, \aap, 533, A109 
\bibitem[Travaglio et al.(1999)]{travaglio99} Travaglio, C., Galli, D., Gallino, R., et al.\ 1999, \apj, 521, 691 
\bibitem[Travaglio et al.(2004)]{travaglio04} Travaglio, C., Gallino, R., Arnone, E., et al.\ 2004, \apj, 601, 864 
\bibitem[Trippella et al.(2016)]{trippella16} Trippella, O., Busso, M., Palmerini, S., Maiorca, E., \& Nucci, M.~C.\ 2016, \apj, 818, 125 
\bibitem[Trippella \& La Cognata(2017)]{TLC17} Trippella, O., \& La Cognata, M.\ 2017, \apj, 837, 41 
\bibitem[Yong et al.(2012)]{yong12} Yong, D., Carney, B.~W., \& Friel, E.~D.\ 2012, \aj, 144, 95 
\bibitem[Zhao et al.(2016)]{zhao16} Zhao, G., Mashonkina, L., Yan, H.~L., et al.\ 2016, \apj, 833, 225 









 \end{thebibliography}
\end{document}